\documentstyle[12pt,psfig,twoside]{article}
\begin{document}\hbadness=10000
\pagenumbering{arabic}
\thispagestyle{empty}
\pagestyle{myheadings}\markboth{J. Letessier, J. Rafelski and A. Tounsi}
{Strange Particles Yields from QGP}
\title{ENERGY DEPENDENCE OF STRANGE PARTICLE\\ YIELDS FROM A QGP-FIREBALL}
\author{$\ $\\
\bf Jean  Letessier$^1$, Johann Rafelski$^{2}$ {\rm and} Ahmed
Tounsi$^1$\\ $\ $\\
$^1$Laboratoire de Physique Th\'eorique et Hautes Energies\thanks{\em
Unit\'e  associ\'ee au CNRS UA 280.}\\
Universit\'e Paris 7, 2 place Jussieu, F--75251 Cedex 05.
\\ \\
$^2$Department of Physics, University of Arizona, Tucson, AZ 85721\\ }
\date{}   
\maketitle \vspace{-10.8cm}{\bf AZPH-TH/95--13}
\hfill {\bf PAR/LPTHE/95--24}\\  \vspace*{11.3cm} 
\begin{abstract}
{\noindent
We explore, as function of the collision energy and stopping in 
relativistic nuclear collisions, the production
yields of strange particles, in particular strange antibaryons,
assuming formation of a deconfined thermal QGP-fireball which 
undergoes a sudden hadronisation. The non-equilibrium freeze-out 
conditions are established and strange antibaryon excitation 
functions are shown to have characteristic features that should 
allow to discriminate between the QGP hypothesis and other 
reaction scenarios.
}\end{abstract}\vfill

\noindent PACS numbers: 25.75.+r, 12.38.Mh, 24.85.+p \hfill{May/June 1995}\\
\newpage 
\section{Introduction}
 
In relativistic nuclear (A--A) collisions we search for a macroscopic, 
deconfined space time region in which soft quark and gluon degrees of
freedom are present, the so-called quark-gluon plasma (QGP) --- `plasma'
since it involves freely moving color charges within the limited reaction
volume. Existence of this phase of 
matter seems to be an inescapable consequence of our theoretical
knowledge about the fundamental hadronic interactions, qualitatively
rooted in  quantum chromo-dynamics (QCD). Nevertheless, it is important
that we can demonstrate QGP formation experimentally, confirming the
theoretical paradigm and permitting the experimental study of the strong
interactions vacuum state. 
 
Because of its  rather limited (nuclear size) volume and hence
very short lifespan \cite{Kat92} of about 4 fm/c, the study of the
QGP phase is a formidable task.
Moreover it is far from clear that this new phase can be formed and
studied with the presently available nuclear  beams, up to 200A GeV
energy at CERN-SPS. However, we have shown in our recent
work \cite{analyze,analyzea} that strange particle production results
point  towards formation of the deconfined phase. In this paper we develop
in full the method that involves as the observable strange
antibaryons \cite{antib,RD87}. Our primary interest is oriented towards
(multiply) strange antibaryons for a number of reasons: the conventional
backgrounds are small since the multi-step processes required occur
relatively rarely in $p$--$p$ interactions. Consequently, the relatively
high production rate of  multiply strange antibaryons in A--A reactions,
and the central (in rapidity) spectral distribution, are indicative of a
`collective' formation mechanism. In the QGP reaction picture it is the
ready made high density of (anti) strange quarks which leads to highly
anomalous yields of multiply strange particles. 

The use of specific hadronic particles in the search for QGP is only 
possible if the evolution  of the final state is such that information
about key properties of the primordial source is retained during the  
formation and evolution of the final state hadrons. This is observed
experimentally indicating that there is no opportunity for the strange
antibaryons emerging from deconfined state  to re-equilibrate (annihilate) 
\cite{analyze,gammas}, thus implying a sudden transition into the final
hadronic particles \cite{RD87,gammas}.
Of course it is far from certain today that QGP is formed already in the
nuclear collisions below $\sqrt{s}=10$A GeV. Several studies addressed
recently the question  \cite{analyze,Let92,Cley93,CR93,Let94a} how the
current data can distinguish between the formation of a central fireball
consisting of a dense and very hot hadron (resonance) gas and the
creation of a transient deconfined state such as the quark-gluon  plasma
(QGP) which subsequently hadronises. We note that even if this final
state hadronisation involves a full re-equilibration, then the memory of
the QGP should in some circumstance not be fully lost, since the deconfined
state is often richer in entropy than the confined
state \cite{Let92,entropy,Divonne,MG}, which becomes visible as particle excess in
the final state. Today, based on diverse results obtained at 200A GeV 
\cite{analyze} it appears that we have a very favorable situation with
the observable particles very much representing the expected properties
of the primordial QGP state, while the results from BNL-AGS at 10-15A
GeV  are indicating that if QGP was formed, it has undergone a full
re-equilibration transition \cite{BNL-AGSthermal}. Consequently, somewhere
between the two energy regimes a major change in the reaction mechanism
occurs. It appears to be of considerable importance to find in which
energy range this change occurs and how other observables vary with
energy in this changeover region.
 
Since the QGP picture works so well at 200A GeV, it is natural to expect
that its formation is also occurring at somewhat lower energies.
Therefore, within a  dynamical collision model, using the equations of 
state (EoS) of QGP, we study in this paper
how the thermal and chemical properties of the fireball of the dense
hadronic matter change with energy. We predict here the production of
strange particles as function of energy. Our primary goal is to find how
the strange particle abundance in the final state is related to the
conditions prevailing in the early moments of nuclear collisions. We
develop an approach in which the collision force is counteracted by the
thermal radiative  pressure. The tacit assumption that we make is that
the thermal equilibrium of quarks and gluons arises rapidly as a result
of the primary interactions. This introduces without an explicit
mechanism almost 70\% of the final state  entropy  \cite{cool} --- this is 
the excess QGP entropy  \cite{entropy} visible as the excess multiplicity
of produced particles. Thus in our approach we do not resolve the
important issue, what is the energy and more generally the conditions
required in the collision for the formation of the deconfined state. On
the other hand, the principal virtue of the thermal model framework
adopted here is that the spectra and particle abundances can be
described in terms of a few parameters which have simple physical
meaning.

Aside of the rather demanding multi-strange antibaryon signature of the
QGP, there is the more readily available, but less specific overall
abundance of strangeness. Strange particle signatures of dense
baryon rich hadronic matter have been theoretically considered for nearly
two decades  \cite{KMR86}. Because of the intrinsic difficulties related
to the detection of strange particles within the large multiplicity of
hadrons produced in relativistic nuclear collisions  only recently
results have been presented which allow a test of the different
theoretical models  (for a recent review of available results, see 
Ref.\, \cite{AIP}). Because strangeness production processes constitute a
bottleneck in chemical equilibration of conventional confined strongly
interacting matter  \cite{sprodQGP,Zim81,sprodHG,KMR86}, strangeness 
production in QGP has been widely studied 
 \cite{antib,sprodQGP,Zim81,sprodQGPa}. In the theoretical
description several microscopic features combine to produce this
enhancement: there is a high quark-gluon  density and the
production threshold of strange quark pairs in QGP is below the mean
energy (temperature) of the constituents present. The computed
strangeness relaxation time constant $\tau_{\rm s}\leq 6$ fm/c towards
the attainment of the strange phase space equilibrium abundance in the
QGP phase is of the magnitude similar to the estimated lifespan
$\tau_{\rm QGP}\simeq R/c$ of the dense QGP state, where $R$ is the
characteristic size parameter of the QGP phase. Consequently, even if the
magnitude of both these time constants is still rather uncertain, their
similarity is of considerable importance for the exploration of the QGP
phase: varying the size of the colliding nuclei, or the mean impact
parameter in collision, we can change the size $R$ of the interaction
region and thus change the lifespan of the dense state. Since $\tau_{\rm
s}$ depends primarily on the temperature (mean glue energy), we could 
explore how this impacts the strangeness chemical
equilibration and compare with the theoretical predictions. 
 
A crucial confirmation of the reaction picture would arise if the changes
observed in the chemical equilibration of strangeness  would be equally
attainable when the energy content in collision is varied. More
importantly, this could be done at the maximum volume available, thus
assuring that unwanted dynamical variations such as changes in
longitudinal flow (transparency) with changes in impact parameter were
minimal, while the highest accessible particle and energy density is
explored in each case. For the strange antibaryon signature this approach
is much preferred, as the abundance of particles produced, and their
centrality remains assured. Since the (enhanced) production of
(multi)strange antibaryons is widely considered to be more specifically
related to the  deconfinement than  the strangeness enhancement in
general, it would be of considerable importance to obtain data on
(multi)strange antibaryon production as function of energy. 
 
Within our approach one of key tacit assumptions is that there is
formation at central rapidity of an energy and baryon-rich, 
fireball. We believe that the  pattern of diverse  particle production
processes confirms the view that even the 200A GeV  collisions are
leading to the formation of a domain of space-time in which not only
energy, but also considerable baryon number is present. For example, the
rapidity distribution  of  $\Lambda$ and $\overline{\Lambda}$ as measured
by the NA35 experiment  \cite{NA35AIP} affirms such a picture of the A--A
reactions. These results show that the S--Ag/W/Pb and even the S--S 
collisions at 200A GeV are far from the limit of baryon-transparency in 
which the valence quarks of projectile and target are  presumed to 
leave the central  rapidity region  \cite{BJO83}. Relatively large 
stopping behavior at 200A GeV  could
be expected: assuming that a  valence quark carries,  in a typical case,
about 1/6 of the energy of a nucleon, and placing the observer in the  CM
frame, we find that we would have two 1.5 GeV partons colliding at CERN-SPS
energies ---  at BNL-AGS energies (10-15A GeV ) a head-on collision
interpreted in this picture would occur between partons of 0.4~GeV. 
Moreover, we have in each hadron a rather wide
distribution of parton energies,  and thus many  collisions between sea
and valence partons are occurring at still considerably lower energies. 
Just a few elastic or inelastic interactions will suffice to stop many 
partons in the central fireball. It is thus not
surprising  that in the particle spectra there is little room, if any,
for the separation in rapidity of energy and baryon number. 

In this consideration it is important to
remember that  the more easily analyzed and understood symmetric A--A
collisions at small impact parameters the surface (corona) nucleons  will
undergo interactions which resembles the normal $p$--$p$ collision
environment. These $p$--$p$ like components
should not be confused with the more specific elements of the 
A--A interactions --- strange particles are a good tag on these specific
interactions, since strangeness  production in A--A collisions is
enhanced compared to $p$--$p$ and $p$--A based expectations, and because 
there are a number of particles (such as $\overline{\Lambda}$) which are
rather rarely produced in  $p$--$p$ collisions, in contrast to
 A--A interactions, and which show, very strongly in these
interactions, features we would interpret as originating in the central
fireball.
 
Our paper is organized as follows: in next section \ref{Init} we explore
the conditions reached in the early deconfined QGP fireball formed e.g.
in  the CERN-SPS collisions. We then compare in section \ref{results} the
results we obtain with those derived for the S--W/Pb at CERN-SPS and
Au--Au interactions at BNL-AGS. We obtain an excellent agreement with all
the  available data. We also make predictions about
strange particle production for the forthcoming round of experiments 
and about the expected particle yield behavior as 
the energy of the collision is varied. 
 
\section{Properties of the QGP State}\label{Init}
\subsection{The Thermal Approach}
Application of the thermal model to hadronic processes implies that on 
the prevailing time scale of the interaction, thermalization of the 
relevant degrees of freedom is rapid. This means that thermal features 
are seen in the particle production processes \cite{HRblack} which are not 
rooted directly in the microscopic features of strong interactions (QCD). 
We can expect that the leading  particle and flow effects 
diminish in their significance as the size of the interaction region 
increase. Thus soft hadronic physics in the A--A case should be easier 
to understand. The original motivation to study these large nuclei in 
highly relativistic collisions was just the hope that we will 
unravel the soft hadronic physics which has remained 
in veil when studied with the  $p$--$p$ interactions. 
 
It is convenient to incorporate in the model development the particle 
spectra as function of the rapidity  $y$ and transverse mass 
$m_\bot$ of a particle
\begin{equation}
y= {1\over 2}\ln\left( {{E+p_z}\over {E-p_z}}\right)\,,\qquad
E=m_\bot \cosh y\,,\qquad m_\bot=\sqrt{m^2+p_\bot^2}\,,
\end{equation}
where `$\bot$' is perpendicular to the collision axis `$z$'. While  $m_\bot$
is invariant under Lorenz transformations along the collision axis, the 
particle rapidity  $y$ is additive, that is it changes by the constant
value of the transformation for all particles. This allows to choose the 
suitable (CM --- center of momentum) reference frame characterized by its 
rapidity $y_{\rm CM}$ for the study of the particle spectra. 
 
The `thermal'  hypothesis presupposes that the
shape of these spectra is sufficiently similar to allow a reduction of
all  data  to just one `Boltzmann' spectral  shape centered around the
fireball rapidity $y_{\rm CM}$:
\begin{equation}\label{spectra}
{dN\over d^3p}=N_i e^{-E^{(i)}/ T}=
     N_i e^{-\cosh (y-y_{\rm CM})\,{m^i_\bot/ T}}\,.
\end{equation}
The parameters of each particle distribution \cite{RRD92}
 include the inverse slope
$T$ (`temperature') of the $m_\bot$ distribution, centered around the
$y_{\rm CM}$. In particular the temperature $T$ derived from the
transverse mass distribution  should be common for all 
types of particles produced by the same thermal mechanism. This relates 
many particle spectra to each other and reduces  the number of
observables considerably. Similarly, $y_{\rm CM}$ should be also the
same for different particles originating in the same thermal source. 
In addition, for each particle there is the normalization 
constant $N_i$. We will discuss below how these constants are determined 
by a chemical equilibrium parameters, aside of the volume $V$. 
Let us note in passing that the presence of a common inverse slope 
parameter for different particle spectra, which is different from 
the one found in $p$--$p$ collisions is well established experimentally. 
For example, for the 200A GeV S--W/Pb interactions the inverse slope 
parameters of strange baryons and antibaryons \cite{WA85AIP} and for 
example of high $m_\bot$ neutral mesons  \cite{WA80spec} are all 
consistent with $T=232$ MeV.

Since a wealth of experimental data can be described with just a few 
model parameters, this leaves within the thermal model a considerable 
predictive power and a strong check of the internal consistency of thermal
approach. Specifically, in the directly hadronising  off-equilibrium 
QGP-fireball model  \cite{analyze} there are 4 parameters
(aside of $T$ and $y_{\rm CM}$) characterizing all particle 
spectra:  two 
fugacities $\lambda_{\rm q},\,\lambda_{\rm s}$, and two particle 
abundance non-equilibrium parameters, the strangeness occupancy 
factor  $\gamma_{\rm s}$ and   the ratio $R^{\rm s}_c$ related to
meson and baryon abundances (see section \ref{Counting}).
Moreover, these physical 
parameters can be determined using a dynamical picture  of the collision, 
in which the input is derived from more general qualitative conditions of 
the colliding system, such as the energy content or stopping power.  Thus
the validity of thermal and chemical equilibrium can be conclusively 
tested, comparing the predicted particle  spectra and yields with 
theoretical predictions without the need or capability to modify and 
adapt the description to each new  result which is reported.
 
\subsection{Parameters  of the  Fireball}\label{param}
We now introduce and discuss in qualitative terms the model parameters 
and their temporal evolution during nuclear collision. Practically all
results we will obtain will hinge on the time
evolution scenario we adopt. At the present time the experimental
constraints are not limiting our reaction and evolution picture in a
unique way. The picture we adopt is in qualitative, and as we shall see,
also in quantitative agreement with a number of important experimental
results, and it is in accord with the general properties of the strong
interactions and hadronic structure widely known and accepted today. 
\subsubsection{Temporal Evolution of the Fireball}\label{temp}
We suppose that the relevant time development stages of the relativistic
nuclear collision comprise:
\begin{itemize}
\item[\it 1.] The pre-thermal stage lasting perhaps 0.2--0.4 fm/c,
during which the thermalization of the initial quark-gluon distributions
occur. During this time most of the entropy obtained in the collision
must be created by mechanisms that are not yet understood --- this is
also alluded to as the period of de-coherence of the quantum collision
system. Our lack of understanding of this stage will not impact our
results, as the reason that we lack in understanding is that the
hadronic interactions erase the memory of this primordial stage.
\item[\it 2.] The subsequent inter-penetration of the projectile and the
target lasting about $\sim 1$ fm/c, probably also corresponding to the
time required to reach chemical equilibrium of gluons $G$ and light
non-strange quarks $q=u,\,d$\,. 
\item[\it 3.]  A third time period (3--5 fm/c) during which the
production and chemical equilibration of strange quarks takes place. During
this stage many of the physical observables studied here will be
initiated.
\item[\it 4.] Hadronisation of the deconfined state ensues: it is
believed that the fireball first expands at constant specific entropy
per baryon, and that during this evolution or at its end it decomposes
into the final state hadrons, under certain conditions in a (explosive)
process that does not allow for re-equilibration of the final state
particles.  
\end{itemize}
In the sudden hadronisation picture of the QGP fireball suggested
by certain features seen in the analysis  of the strange antibaryon 
abundances for the 200A GeV  nuclear collision 
data  \cite{analyze,analyzea,gammas}, the hadronic observables 
we study are not overly sensitive to the details of
stage {\it 4}. Akin to the processes of direct emission, 
in which strange particles are made in recombination--fragmentation 
processes \cite{RD87}, the chemical conditions prevailing in the
deconfined phase  are  determining  many relative final particle yields. 
If the hadronisation occurs as suggested by recent lattice results 
\cite{lattice} at a relatively low temperature (e.g. 150 MeV), 
the total meson abundance which is governed by the entropy content at
freeze-out of the particles, is found about 100\% above the hadronic 
gas equilibrium expectations \cite{entropy}. This is consistent with 
the source of these particles being the QGP  \cite{entropy,analyze}. 
The freeze-out entropy originates at early time in collision since
aside of strangeness production which is responsible for about 10\%
additional entropy there is no significant entropy production after
the initial state has occurred \cite{entropy}. 
Similarly, the relatively small thermal abundance of baryons must 
be enhanced by  the factor 5 in order to maintain a ratio $R^{\rm s}_c
=0.4$  \cite{analyze}.

On the other hand the  experimental results obtained at 15A GeV are 
consistent with the thermal equilibrium hadronic gas state expectations
 \cite{BNL-AGSthermal}. In the event that the source of these particles
should indeed be a QGP fireball, a slow re-equilibration  transition 
should be envisaged here. In this scenario the details of hadronisation 
mechanisms are of even lesser relevance  since the equilibrium state is 
reached among the final hadron gas particles.

\subsubsection{Temperature}
During the initial contact of the two nuclei, as
soon as stage {\it 1} is reached, it 
is likely that the most extreme conditions 
(highest temperature) prevails. Subsequently,
temperature decreases rapidly, primarily due to two mechanisms: there is
rapid quark and gluon production, leading towards the chemical
equilibration, and second, the fireball expansion/emission  cooling
occurs. We will account for these two effects balancing energy  per
baryon or entropy per baryon of the fireball as appropriate for the 
evolution stage. In this way, we implement during the stage {\it 2} 
a one dimensional, hydrodynamical evolution of the dense deconfined
hadronic matter, in addition allowing for the approach to chemical 
equilibrium. We encounter a considerable drop in temperature between the 
initial stage {\it 1} and the final state {\it 4} freeze-out point, 
but the entropy content which determines
the final particle multiplicities remains nearly constant: aside of the
initial state entropy formation all the additional increase is due to the
formation of the strangeness flavor.
 
The different stages of the evolution are characterized by the
temperatures:
\begin{center}
\begin{tabular}{cl}
$T_{\rm th}$ & temperature associated with initial thermal equilibrium,\\

$\downarrow$ & {\it production of} $q,\ {\bar q},\ G$, {\it expansion;}
\\

$T_{\rm ch}$ & temperature of chemical equilibrium for non-strange 
quarks and gluons, \\

$\downarrow$ & {\it production of} $s,\ {\bar s}$ {\it quarks and
fireball expansion;} \\

$T_{\rm 0}\ $ &  temperature of maximal chemical equilibrium: 
`visible' temperature,\\

$\downarrow$ &  {\it fireball expansion/particle radiation}; \\

$T_{\rm f, s}$ & temperature at freeze-out for non-strange and 
strange particles, 
\end{tabular}
\end{center}
with obviously $T_{\rm th} > T_{\rm ch} > T_{\rm 0} \ge T_{\rm f}$. 
 
In the transverse mass spectra of strange (anti)baryons a temperature  
$T_\bot$ is found. If the final state particles emerge 
directly without re-equilibration from the fireball \cite{RD87,gammas}, 
this  observed temperature ($T_\bot=232\pm5$ in S--A
collisions at 200A GeV) in the particle spectra would be closely 
related to the  full chemical equilibration temperature $T_{\rm 0}\ $: 
subsequent to the establishment of the conditions at $T_{\rm 0}\ $ we 
have either directly the emission of particles and thus we have  
$T_{\bot}\le T_0$, or there is collective,  (so called transverse) 
radial flow in the hot matter, in which fraction  of the thermal 
energy is converted into the  flow energy.  When  the final state 
particles emerge from the flowing surface, they are  blue-shifted by  the
flow velocity. This Doppler shift effect restores the high apparent 
$T_0$  in high $m_\bot$ particle spectra \cite{ULI93}:
\begin{eqnarray} 
T_{\bot}\le T_0\equiv\sqrt{1+v_{\rm f}\over 1-v_{\rm f}} T_{\rm
f}\,,\qquad
T_{\bot}\le T_0\equiv\sqrt{1+v_{\rm s}\over 1-v_{\rm s}} T_{\rm s}\,; 
\label{doppler} \end{eqnarray}
--- strange and non-strange particles may not originate in
exactly the same condition and hence we introduced here the 
option to have different freeze-out temperature. In view
of the smaller strange particle cross sections we expect  $T_{\rm
s}\ge T_{\rm f}$.
 
As suggested in Eq.\,(\ref{doppler}), the temperature $T_0$ at the
onset of  flow is close to the inverse slope 
spectral temperature $T_\bot$ implied by the high $m_\bot$
(strange antibaryon) particle spectra. Despite our ignorance of the 
freeze-out mechanisms and conditions, we believe that the uncertainty  in
the value of the initial temperature as derived from the value of
$T_{\bot}$ in  Eq.\,\ref{doppler} is not large. If QGP phase is directly
dissociating by particle emission, this is trivially so. If there were to
be substantial flow, one can assume some temperature $T_{\rm 0}$, and 
given the EoS, compute accurately the hydrodynamic radial
expansion \cite{Kat92}; at the high $m_\bot\simeq 2$ GeV considered here, 
using Eq.\,(\ref{doppler}) we find that 
the inverse slope $T_\bot$ of the particle  distributions is equal or a
bit smaller than $T_{\rm 0}$. 
 
\subsubsection{Fireball Rapidity}
The fireball is created in central symmetric collisions at  the 
CM-rapidity of the N--N system, which is for relativistic systems 
just is 1/2 of the projectile rapidity. For asymmetric collisions such as 
S--Au/W/Pb the CM rapidity depends on the ratio of the participating
masses of the projectile $A_{\rm P}$ and target $A_{\rm T}$ nuclei. 
The center of momentum frame is  \cite{Sch88} (neglecting small corrections):
\begin{equation}
y_{\rm CM}={y_{\rm P}\over 2}-{1\over 2}\ln {A_{\rm T}\over A_{\rm P}}\,.
\end{equation}
Assuming small impact parameter collisions with a suitable central
trigger,  all projectile nucleons participate while the target
participants $A_{\rm T}$ can be estimated from a geometric `interacting
tube' model \cite{Sch88}. 
This model reproduces well the central value  of rapidity center
 particle  spectra  in the specific case of 200A GeV S--Au/W/Pb
interactions. One in particular finds $y_{\rm CM}=2.6\pm0.1$, the 
uncertainty arising from the impact parameter averaging and variations 
of the surface nucleon participation. Once the central rapidity
is defined, and the ratio of participating projectile and target
masses is known, it is  possible to determine the CM-energy involved in
the interaction. It is from this simple kinematic considerations that 
we derive the energy values presented below in  section \ref{energ}. 
 
\subsubsection{Particle Fugacities}
Among the  chemical (particle abundance) 
parameters there are the well known particle fugacities, which allow to
conserve flavor quantum numbers. Three fugacities are 
introduced since the flavors $u,\,d,$ and $s$ are
separately conserved on the time scale of hadronic collisions and can
only be produced or annihilated in particle-antiparticle pair production
processes\footnote{In many applications it is sufficient to combine the
light quarks into one fugacity 
$\lambda_{\rm q}^2\equiv\lambda_{\rm d}\lambda_{\rm u}\,,\ 
\mu_{\rm q}=({\mu_{\rm u}+\mu_{\rm d}})/2\,.$
The slight asymmetry in the number of $u$ and $d$ quarks is described by
the small quantity $\delta\mu=\mu_{\rm d}-\mu_{\rm u}\,,$
which may be estimated by theoretical considerations \cite{analyze}.}.
 
The fugacity of each hadronic 
particle species is the product of the valence quark fugacities, 
thus, for example, the hyperons have the fugacity 
$\lambda_{\rm Y}=\lambda_{\rm u}\lambda_{\rm d}\lambda_{\rm s}$.
Fugacities are related to the chemical potentials $\mu_i$ by:
\begin{equation}
\lambda_{i} =e^{\mu_{i}/T}\,,\quad 
\lambda_{\bar{\imath}}=\lambda_i^{-1}\qquad 
i={u,\,d,\,s}\, .   \label{lam}
 \end{equation}
Therefore,  the chemical potentials for particles and 
antiparticles are opposite  to each other, provided that there
is complete  chemical equilibrium, and if not, that the deviation from the 
full phase space occupancy is accounted for by  introducing a 
non-equilibrium chemical parameter $\gamma$ (see below). 
 
\subsubsection{Phase Space Occupancy}
Thermal and chemical equilibria are two very different phenomena. In
general, the production of particles is a considerably slower process
than elastic hadronic collisions, and thus even if we assume in our work a
thermal equilibrium scheme, we should not expect the chemical equilibrium to
be present. In addition, there is the relative and absolute chemical
equilibrium. In
the former, the particle abundances are in relative equilibrium with each
other, in the latter the total particle yields are just filling the full
available phase space ---  relative chemical equilibrium is in general
easier to attain than the absolute chemical equilibrium. Our picture of
rapid QGP fireball disintegration in which no equilibration takes place 
in the final state, implies that chemical (abundance) non-equilibrium
features should be present in the final state.  
 
Strangeness flavor provides an interesting example of the above. 
Calculations \cite{sprodQGP,sprodQGPa} of the chemical relaxation 
constant show that in general  it will not fully saturated the 
available phase-space. Therefore, we need to introduce the associated  
off-equilibrium parameter $\gamma_{\rm s}$.  Since the thermal equilibrium 
is, as discussed above, established within a considerably shorter time
scale than the (absolute) chemical equilibration of strangeness, we can
characterize the saturation of the strangeness phase space by an average
over the momentum distribution:
\begin{equation}\label{gamth}
\gamma_{\rm s}(t)\equiv {
           \int d^3\!p\,d^3\!x\,n_{\rm s}(\vec p,\vec x;t)\over 
     \int d^3\!p\,d^3\!xn_{\rm s}^{\infty}(\vec p,\vec x)}\ ,
\end{equation}
where $n_{\rm s}^\infty$ is the equilibrium particle density. The factor
$\gamma_{\rm s}$ thus enters the phase space momentum  Boltzmann
distribution as a multiplicative factor, 
and with the $\vec x$ dependence contained solely
in the statistical parameters we have:
\begin{equation}
n_{\rm s}(\vec p,\vec x;t)\simeq\gamma_{\rm s}(t)n_{\rm s}^\infty
     (\vec p;T(\vec x,t),\mu_{\rm s}(\vec x,t))\,.
\end{equation}
 
A further refinement should be noted: when the quantum aspects of the 
particle distributions are incorporated and the maximum entropy state  of
an isolated physical system (closed system) is obtained, the absolute 
chemical equilibrium coefficients $\gamma_i$ enter as  multiplicative
coefficients in front of the Boltzmann factor within the quantum 
Bose/Fermi distribution  \cite{cool}:
 \begin{equation}\label{qdist}
n^{\rm B,F}_i=
     {1\over{\gamma^{-1}_i\lambda_i{\rm e}^{\beta\epsilon_i}\mp 1}}\,,
\end{equation}
along with the fugacity  factors .
 
The final freeze-out value of $\gamma_{\rm s}$ is a parameter which impacts
the particle ratios in a two fold way: 
there is the simple proportionality (in the Boltzmann limit)
of the particle yields to $\gamma_{\rm s}^n$, --- where $n$ is the 
number of strange quarks contained in the hadron considered,  
and furthermore, the value of
$\gamma_{\rm s}$ impacts the cooling that the fireball undergoes to
produce strangeness. Unexpectedly, it is rather straightforward to 
extract from the strange antibaryon experimental particle yields
 \cite{gammas,analyze,analyzea,Let94a} the value of $\gamma_{\rm s}$. 
Roughly speaking, $\gamma_{\rm s}$ makes its appearance in all
particle ratios in which we compare the abundances involving different
strangeness content.  The measurement of $\gamma_{\rm s}$ is thus 
an important step towards the understanding of
the behavior of highly excited hadronic phases: $\gamma_{\rm s}$ can
be studied varying a number of parameters of the collision, such as the
volume occupied by the fireball (varying size of the colliding nuclei and
impact parameter), the trigger condition (e.g. the inelasticity), the
energy of colliding nuclei when searching for the threshold energy of
abundant strangeness formation.
 
The theoretical dynamical model to investigate $\gamma_{\rm s}(t)$ has
been developed to considerable detail  \cite{KMR86b,gamevolv} 
 --- it arises from a standard population evolution equation and its 
interplay with the expansion-dilution-dissociation of the fireball. 
For a QGP state with its fast gluonic strangeness production even the
natural short fireball lifetime of only a few fm/c should be nearly
sufficient to reach values of $\gamma_{\rm s}\simeq 1$. Detailed
balance assures that the production and annihilation processes are
balancing each other as $\gamma_{\rm s}\to 1$\,. At large times the
approach to equilibrium takes the form  
\begin{equation}\label{gammaap}
1-\gamma_{\rm s}\to 
     e^{-t/\tau_{\rm s}},\qquad t\gg \tau_{\rm s}\,,\end{equation}
where $\tau_{\rm s}$ is the relaxation time for strangeness equilibration
which can be computed using standard QCD methods. A recent evaluation of
this work   \cite{sprodQGPa} confirmed that the glue fusion processes 
 \cite{sprodQGP}  are  dominating the strangeness production rates
in QGP, with $\tau_{\rm s}\simeq  2$ fm/c for the here 
relevant $T\simeq 250$ MeV temperature range. The experimental results
yield $\gamma_{\rm s}=0.7$--1, imply according to Eq.\,(\ref{gammaap}) a
lifespan of the the plasma state to be greater than 3 fm.
 
In a purely HG fireball, with its expected much longer strangeness
saturation time scale, small values of $\gamma_{\rm s}\sim$ 0.1 like
those extracted from $N$--$N$ collisions should prevail  \cite{sprodHG}. 
A measurement of $\gamma_{\rm s}$ thus in principle provides important
information on the strangeness production time scale and hence a
relatively large value of $\gamma_{\rm s}$ requires some new physics
feature in the structure of the collision fireball, which we like to
associate with QGP formation. 
 
\subsection{QGP Equations of State}\label{EOSQGP}
The partition function of the interacting quark-gluon phase can be
written as:
\begin{equation}\label{ZQGP}
\ln Z^{\rm QGP}=\sum_{i\in \rm QGP} {{g_i(\alpha_s) V}\over{2\pi^2}} \int
\!\pm\ln\left(1\pm \gamma_i\lambda_i e^{-\sqrt{m_i^2(T)+p^2}/T} 
\right) p^2\,dp\,, \end{equation}
where $ i=G,\ q,\ \bar q,\ s,\ \bar s$, with $\lambda_{\bar\imath} =
\lambda_i^{-1}$ and $\gamma_{\bar\imath}=\gamma_i$. We take into account
the QCD interaction between quarks and gluons by allowing for thermal
masses 
\begin{equation}
m_i^2(T)=(m_i^0)^2+(c\,T)^2\,.
\end{equation} 
For the current quark masses we take: 
$$m_{\rm q}^0=5\  {\rm MeV},\quad m_{\rm s}^0=160\  
{\rm MeV},\quad  m_{\rm G}^0=0\,.$$
We have $c^2\propto\alpha_{\rm s}$, $\alpha_s$ being the QCD coupling 
constant. Considering the 
theoretical uncertainty regarding the coefficient $c(\alpha_{\rm s})$, 
we explored two different approaches:
\begin{itemize}
\item[\it 1.] We fix $c=2$, arising for $\alpha_s\sim 1$ 
(the exact value was not of essence),
and allowing for another effect of the QCD-interaction, 
the reduction of the number of effectively available 
degrees of freedom.
\indent We implement the following effective counting of 
gluon and quark degrees
of freedom, motivated by the perturbative QCD formul\ae:
\begin{eqnarray}
g_{\rm G}&=16&\to\quad 
   g_{\rm G}(\alpha_s)=  16\left(1- {15\alpha_s\over 4 \pi}\right)\,,
\nonumber\\
g_{i-{\rm T}}&=6&\to\quad  
 g_{i-{\rm T}}(\alpha_s)=    6\left(1-{50\alpha_s\over 21\pi}\right)\,,
\label{eq12}\\ 
g_{i-{\rm B}}&=6&\to\quad  
  g_{i-{\rm B}}(\alpha_s)=   6\left(1-2{\alpha_s\over \pi}\right)\,,
\nonumber
\end{eqnarray}
where $i=u,\ d,\ s$.  
In Eq.\,(\ref{eq12}) two factors are needed for quarks:  the factor 
$g_{i-{\rm T}}$ 
controls the expression when all chemical potentials vanish 
(the $T^4$ term in the partition function for massless quarks) 
while $g_{i-{\rm B}}$ is taken as coefficient of the
additional terms which arise in presence of chemical potentials.
We took $\alpha_s=0.6$ which turned out to be the value best suited 
for the experimental data points\footnote{This value arises 
at $T=2.2\,T_c$ in the fit \cite{Kam94,Kam95} to lattice gauge data 
without quarks.}.  
\item[\it 2.] A fit proposed by K\"ampfer { et al.} \cite{Kam94,Kam95}
made to the lattice data for the SU(3) gauge sector and extended to quarks
\begin{eqnarray}
c^2&=&3.8\alpha_s^{\rm m}(T)\,;\\
\alpha_s^{\rm m}(T)&=&{{4\pi}\over {9\ln (T/T_c+0.023)^2}}\,,
\end{eqnarray}
where the form of $\alpha_s^{\rm m}$ is motivated by $\alpha_s(q^2)$.
 The critical temperature of the QGP transition 
to confined phase is chosen at $T_c=150$ MeV  and the infrared cutoff at 
$ 0.023 T_c$. 
\end{itemize}
The main difference between these two approaches 
is how the quark and gluon degrees of freedom are
suppressed  near to the critical temperature. 
In both cases these phenomenologically motivated 
procedures are a bold extension of the established 
perturbative and/or lattice QCD results. 
 
\begin{figure}[tb]
\vspace*{-1cm}
\centerline{\hspace*{-1.2cm}
\psfig{width=18cm,figure=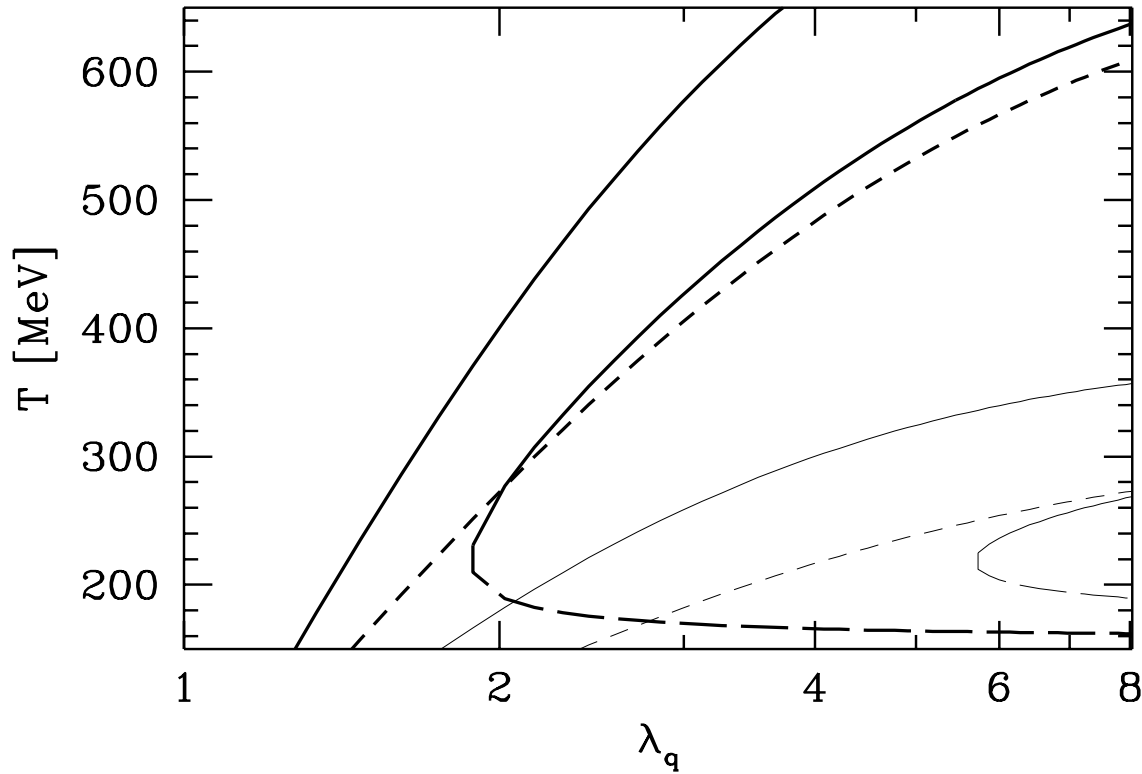}
}
\vspace*{-0.5cm}
\caption{\protect\small 
Properties of different QGP-EoS: constraint between 
temperature $T$ and light quark fugacity $\lambda_{\rm q}$ for a given energy 
content per baryon $E/B=4.3$ (thin lines) and $ 9.6 $ GeV (thick lines). 
Upper solid line: procedure {\it 1} of EoS (see text), 
lower solid lines joining onto the long-dashed lines (second solution branch):
procedure  {\it 2}  EoS  and short-dotted: free quark-gluon gas.
 \protect\label{figkam}
}
\end{figure}
Given these interaction effects, the Bose/Fermi
distributions for quarks/gluons, Eq.\,(\ref{ZQGP}), 
allow us to obtain in a consistent way any physical 
property of this model EoS of the QGP phase. These 
are shown in Fig.\,\ref{figkam}. We show here in $T$--$\lambda_q$ 
domain lines of fixed energy  per baryon (thick lines: 9.6 GeV, 
thin lines 4.3 GeV) --- solid lines are for the 
procedure {\it 1}, dashed lines are the ideal gas (no interaction) 
lines, long dashed is the second solution branch of the procedure {\it 2}, 
which attaches to the primary solution branch also shown as solid lines. 
Note that the second solution branch is probably an artifact,
as it tangents the critical temperature, where this method is questionable. 
Interestingly, we note that at high $T$ the perturbative result is 
just below the results of procedure {\it 2}, indicating that the 
change in energy density and baryon number density due to thermal 
masses is equal, and thus the ratio is similar to the free gas result. 
There is substantial difference between the procedure {\it 1} and 
the free gas, which originates in different changes of the degrees 
of freedom due to perturbative interactions in the dominant terms 
of energy density and baryon density.

This representation of the ratio of two densities amplifies the impact 
of interactions and shows how these are incorporated in the 
different approaches. We see that while in the procedure {\it 2} 
fitted to lattice data the treatment of the energy density is particularly 
accurate for baryon-less plasma, for finite baryon density of 
interest here the results are of definitively questionable validity. 
In particular, 
near to the critical temperature (taken here to be $T_c=150$ MeV) 
the large thermal mass of quarks eliminates the baryon density 
effectively and thus it is not even possible to find for a given ratio 
of energy density to baryon density a  reasonable solution for  $T\sim T_c$\,
while for `high' $T$ essentially the free quark-gluon gas result is 
reproduced.
 
Since our present work is primarily concerned with the baryon 
density dependence of the diverse observables near to $T_c$, 
presumably $T_c<T<2T_c$ and $\mu_q\simeq T$,
it is clear that we cannot use the model {\it 2} without prior 
major theoretical improvements applicable to high baryon density domain,
such as is the inclusion of the chemical 
potential dependence in $\alpha_s$ and in the thermal masses, 
and for fermions, the effect of Fermi blocking on the polarization 
function. We thus assume the procedure {\it 1} with parameters 
as described above. 

\subsection{Energy Content in the Fireball}\label{energ}
\begin{figure}[tb]
\vspace*{-1cm}
\centerline{\hspace*{-1.7cm}
\psfig{width=18cm,figure=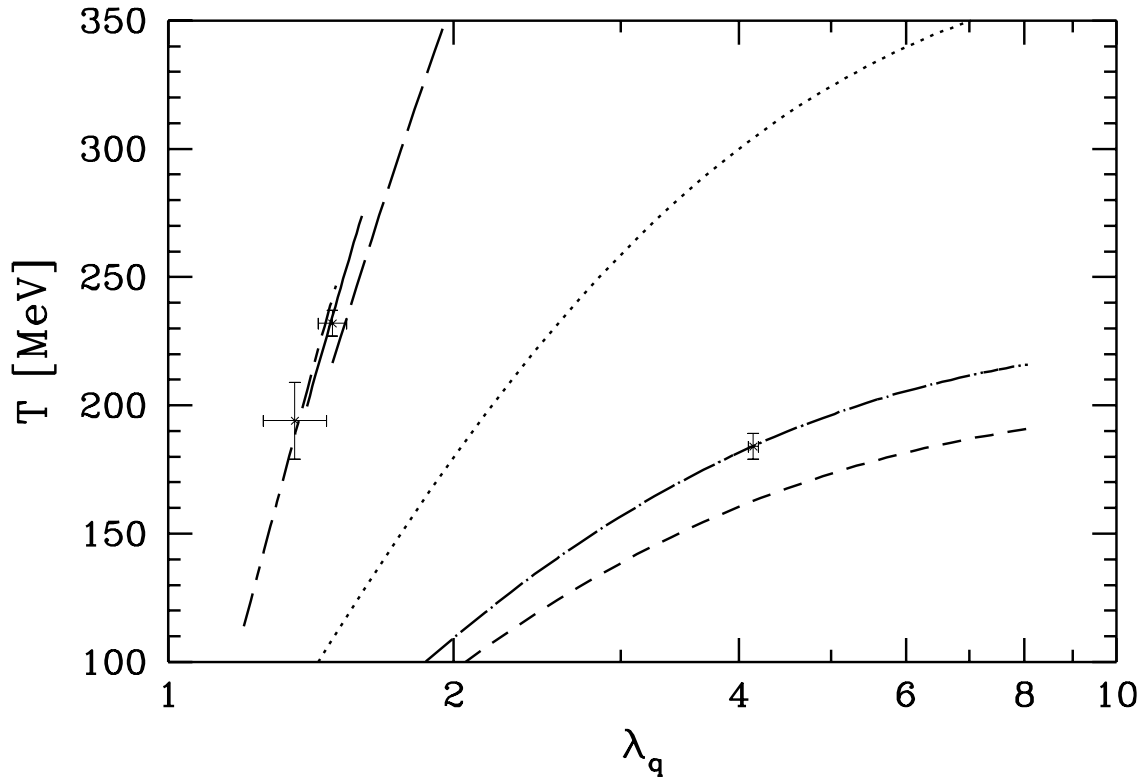}
}
\vspace*{-0.5cm}
\caption{\protect\small 
QGP-EoS according to procedure {\it 1} constraint between temperature 
$T$ and light quark
fugacity $\lambda_{\rm q}$ for a given fireball energy content per baryon 
$E/B$ appropriate for the BNL-AGS and CERN-SPS collision systems. 
Left to right: 2.3 (Au--Au), 2.6 (Si--Au),
4.3 (A--A), 8.6 (Pb--Pb), 8.8 (S--PB/W) and 
9.6 (S--S) GeV. See text for a discussion of experimental
points. 
 \protect\label{energy}
}
\end{figure}

We now commence the evaluation of the initial conditions
reached in the collision: an important constraint arises 
from the energy per baryon content in the fireball. The 
considerations of the previous section allow to relate 
a given energy per baryon\footnote{Here $B$ is the baryon number. 
To avoid confusion the bag constant is denoted ${\cal B}$.}
$E/B$ to the statistical
parameters $T$ and $\lambda_{\rm q}$. On the other hand, 
the collision energy gives
\begin{equation} \label{ECM}
{E\over B}= {\eta_{\rm E}{E_{\rm CM}}\over {\eta_{\rm B}A_{\rm part}}}
\simeq {E_{\rm CM}\over A_{\rm part}},
\end{equation}
where $A_{\rm part}$ is the number of nucleons participating in the
reaction. The last equality follows when the stopping fractions are 
equal --- the experimental particle spectra we are addressing here, 
and in particular the visible presence of baryons in the central 
rapidity region, are implying
that this is a reasonable assumption for the current experimental
domain. In consequence, the energy per baryon in the fireball is to be
taken as being equal to the kinematic energy available in the collision.
In the current experiments we have the following kinematic energy content:
\begin{center}
\begin{tabular}{ll}
Au--Au  at 10.5A GeV  \quad & $\to$\quad $E/B = 2.3$ GeV ,\\ 
Si--Au at 14.6A GeV  & $\to$\quad  $E/B=2.6$ GeV, \\
A--A at 40A GeV  & $\to$\quad  $E/B=4.3$ GeV, \\ 
Pb--Pb at 158A GeV   &  $\to$\quad  $E/B =8.6$ GeV,\\
S--W/Pb at 200A GeV   &  $\to$\quad  $E/B =8.8$ GeV ,  \\
S--S at 200A GeV  & $\to$\quad  $E/B= 9.6$ GeV ,\\
\end{tabular}
\end{center} 
Note that above we assumed collision with the geometric target tube of
matter when the projectile is smaller than the target, 
see section \ref{param}.
The specific energy content $E/B$, given EoS, establishes a
constraint between the thermal parameters. In Fig.\,\ref{energy} we show
in the $T$--$\lambda_{\rm q}$ plane  the lines corresponding to the
constraint on the QGP-EoS according to procedure {\it 1}, 
 see section \ref{EOSQGP},
arising from fixing the energy per baryon at kinematic value listed above
(rising from bottom right to left). In the middle we show 
the lowest CERN-SPS accessible energy, 4.3 GeV, which bridges the  
current CERN-SPS 
domain  shown to the left to the  BNL region on the lower right.
The experimental crosses show the values of  $\lambda_{\rm q}$ arising  
in our data analysis  \cite{analyze,analyzea,BNL-AGSthermal}, combined 
with the inverse slopes temperatures, see section \ref{param},  
extracted from transverse mass particle spectra. The fact that the 
experimental results 
fall on the lines shown in  Fig.\,\ref{energy} is primarily due to 
the choice $\alpha_s=0.6$ --- as this is the usual value in this 
regime of energy it
implies for a QGP fireball EoS hypothesis that the assumption that 
stopping of energy and baryon number is similar deserves
further consideration.

\subsection{Pressure Balance}\label{presbalan}
We have seen that the QGP energy per baryon constraint between 
$\lambda_{\rm q}$  and $T$ (see  Fig.\,\ref{energy}) allows 
for a good agreement with experiment \cite{eb}. There remains 
the issue what physical constraint or principle
determines which of the possible pair of $T,\lambda_{\rm q}$ values 
along the individual curves in Fig.\,\ref{energy} is experimentally 
recorded by the cross shown. We have explored the properties of the
QGP phase along these lines of constant energy per baryon and have 
noticed that with increasing $T$ the pressure in the QGP phase increases,
and that the experimental points coincide  with the dynamical pressure 
generated in the collision.  This confirms the intuitive idea that the 
initial conditions reached in the central fireball arise from the 
equilibrium between the fireball internal thermal and external 
compression pressure.

This condition takes the form \cite{init}: 
\begin{equation}\label{Pbal}
P_{\rm th}(T,\lambda_i,\gamma_i)=P_{\rm dyn}+P_{\rm vac}\,.
\end{equation}
The thermal pressure follows in usual way from the partition function
\begin{equation}\label{Ptherm}
P_{\rm th}={T/V}\ln Z\,,\end{equation} where aside of the temperature
$T$, we encounter the different (quark and gluon) fugacities $\lambda_i$
and the chemical saturation factors $\gamma_i$ for each particle. For the
vacuum pressure we will use:
\begin{equation}\label{Pvac}
P_{\rm vac}\equiv{\cal B}\simeq0.1 \mbox{GeV/fm}^3\,.
\end{equation}
 
The pressure due to kinetic motion follows from well-established
principles, and can be directly inferred from  the pressure
tensor \cite{deG80}:
\begin{equation}\label{Tij}
T^{ij}(x)=\int\! p^iu^j f(x,p){\rm d}^3\!p\,,\quad i,j=1,2,3\,.
\end{equation}
We take for the phase-space distribution of 
colliding projectile and target nuclei
\begin{equation}\label{fxp}
f_{\rm P,T}(x,p)=\rho_{\rm P,T}(x)\delta^3(\vec p\pm\vec p_{\rm CM})\,,
\end{equation}
and hence in Eq.\,(\ref{Tij}) $u^j=\pm p^j_{\rm CM}/E_{\rm CM}$. 
We assume that the nuclear density is uniform within the nuclear size, 
$\rho_{0}=0.16$ /fm$^3$.

\begin{figure}[tb]
\vspace*{-1.5cm}
\centerline{\hspace*{-1.7cm}
\psfig{width=18cm,figure=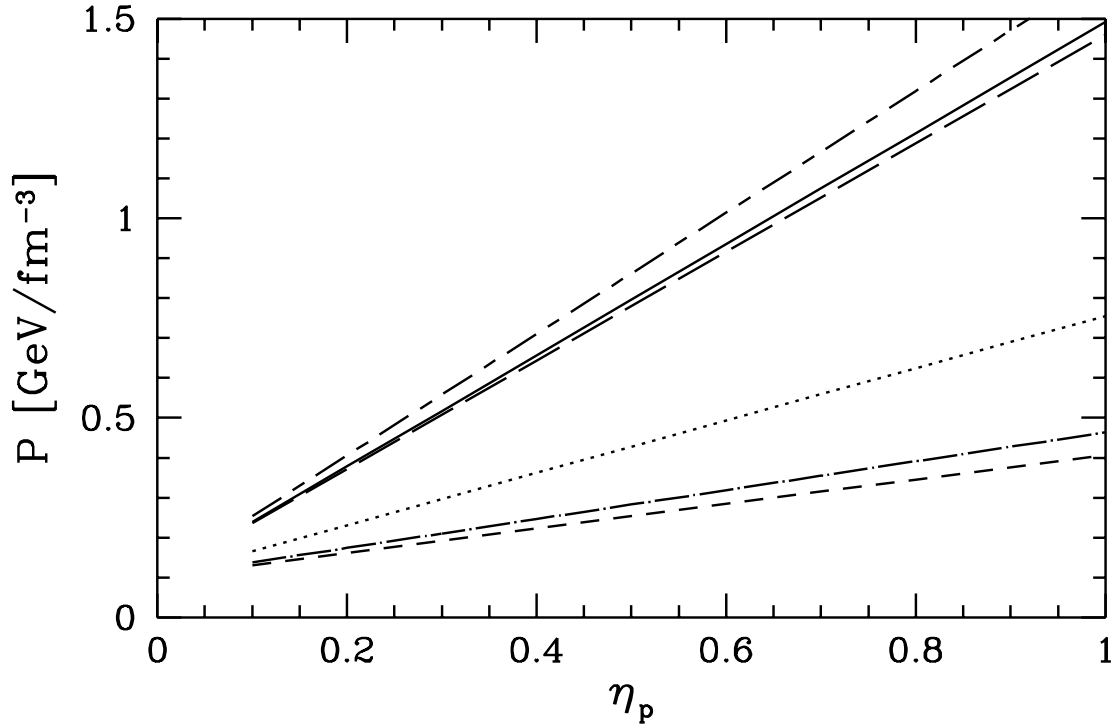}}
\vspace*{-0.5cm}
\caption{\protect\small 
The collision pressure $P$ as function of momentum stopping $\eta_{\rm
p}$ for different values of $E/B$ --- 2.3, 2.6, 4.3, 8.6, 8.8 and
9.6 GeV (from bottom to top, solid line is for 8.8 GeV).
 \protect\label{Pevolution}}
\end{figure}

To obtain the pressure exerted by the flow of colliding matter, we
consider  the pressure component $T^{jj}$, with $j$ being  
the direction of $\vec v_{\rm CM}$. This gives
\begin{equation}\label{Pdyn}
 P_{\rm dyn}=\eta_{\rm p} \rho_{0}\frac{p_{\rm CM}^2}{E_{\rm CM}}\,.
\end{equation}
Here it is understood that the energy $E_{\rm CM}$  and the momentum
$p_{\rm CM}$ are given in the nucleon--nucleon CM frame and $\eta_{\rm
p}$ is the momentum stopping fraction --- only this fraction 
$0\le\eta_{\rm p}\le 1$ of the incident CM momentum can be used by a
particle incident on the central  fireball  (the balance remains in the
un-stopped longitudinal motion) in order to exert dynamical pressure. For
a target transparent to the incoming flow, there would obviously be no
pressure exerted.  In   Fig.\,\ref{Pevolution}  we show, as function
of momentum stopping $\eta_{\rm p}$ the collision pressure $P$. 
For $\eta_{\rm p}\simeq 0.5$ the value of the pressure $P$ is close to 
 the value found in the QGP phase (see table 1). We note that if
indeed in Pb--Pb collisions the momentum stopping increases, the
dynamical pressure can double, causing a considerable increase in the
baryon compression and the associated slow rise of $\lambda_{\rm q}$ as
the size of the collision system increases.
 
\subsection{Final State Statistical Parameters}\label{finals}
We are here primarily interested in determining 
the properties of the (near) chemical equilibrium state of the 
QGP-fireball prior to its disintegration. 
During the time the nuclei interact in collision, we assume
rapid thermal equilibration of the fraction $\eta$ of the kinetic 
energy. The evolution of particle numbers towards chemical 
equilibrium occurs in this stage under the constraint of a given 
constant (compression-)pressure  and energy per baryon. 
Given values of $E_{\rm CM}$ and $\eta_{\rm p}$, as well as taking
$\eta_{\rm E}=\eta_{\rm B}$ we have prescribed a particular energy per
baryon  $E/B$ in the fireball and we can concurrently solve the pressure
equilibrium constraint given by Eq.\,(\ref{Pbal}) to determine the
pressure. Since the pre-equilibrium QGP phase is not observable with 
hadronic probes, much of the detail of the evolution is lost. As we 
find, it suffices to use the energy and pressure 
constraints to determine the 
properties of the fireball at the time the nuclear collision terminates. 
We make here the plausible hypothesis%
\footnote{The finite baryon density and baryon number conservation in the
fireball force onto the system a rather large quark density, which is
there from the beginning and needs not be produced;
gluons are more easily produced than quark pairs and thus presumably
their number catches up with  the quark number by the time the collision
has terminated. Note that in baryon-free central region
environments expected at much higher RHIC/LHC energies, the approach to 
chemical equilibrium can be different.}
that when the collision has 
terminated (at about 1.5 fm/c in the CM frame), the $u$, $d$
quarks and gluons have reached their chemical equilibrium, 
$\gamma_{\rm q}\to 1$, $\gamma_{\rm G}\to 1$. Furthermore,
the strange flavor is still far from equilibrium and we choose
$\gamma_{\rm s}\simeq 0.15$ appropriate for strange quark relaxation 
time 7 times larger than the light quark one \cite{sprodQGPa}. 
Because the QGP phase is strangeness 
neutral we have $\lambda_{\rm s}=1$. The remaining statistical parameters 
$T_{\rm ch}$ and  $\lambda_{\rm q}$ are now fixed by the EoS
and are shown with other interesting properties of the fireball 
(number of gluons per baryon, 
number of light quarks and antiquarks per baryon, number of 
anti-strange quarks per baryon, the pressure in the fireball, 
baryon density and the entropy per baryon)
in the middle section of the table~\ref{bigtable}\,.

\begin{table}[t]
\caption{Properties and evolution of different collision systems.
\protect\label{bigtable}}
\begin{center} 
\begin{tabular}{|c||c|c|c|c|c|c|} 
\hline
Phase&&\multicolumn{5}{|c|}{\phantom{$\displaystyle\frac{E}{B}$}$E/B$ [GeV]
\phantom{$\displaystyle\frac{E}{B}$}}\\\cline{3-7}
\raisebox{1mm}{space}&\raisebox{2mm}{$<\!s-\bar s\!>=0$}
	&2.6&4.3 & 8.8 & 8.6 & 8.6 \\
occupancy&$\lambda_s\equiv 1$ &$\eta=$ 1 
	&$\eta=$ 1&$\eta\!=\! 0.5$&$\!\eta\!=\! 0.75\!$&$\eta=$ 1 \\
&&Au--Au&Pb--Pb&S--Pb&Pb--Pb&Pb--Pb\\
\hline\hline
&$T_{th}$ [GeV]&0.260&0.361&0.410&0.444&0.471\\
$\gamma_q=$ 0.2&$\lambda_q$&9.95&3.76&1.78&1.91&2.00\\
&$n_g/B$&0.20&0.54&1.55&1.36&1.25\\
&$n_q/B$&3.00&3.13&5.12&3.89&3.77\\
$\gamma_g=$ 0.2&$n_{\bar q}/B$&0.00&0.13&2.12&0.89&0.77\\
&$n_{\bar s}/B$&0.02&0.06&0.16&0.14&0.13\\
&$\!P_{th}\!$ [GeV/fm$^3$]&0.46&0.76&0.79&1.12&1.46\\
$\gamma_s=$ 0.03&$\rho_{\rm B}$&3.34&3.30&1.70&2.44&3.18\\
&$S/B$&11.8&18.8&40.0&35.8&33.4\\
\hline
&$T_{ch}$ [GeV]&0.212&0.263&0.280&0.304&0.324\\
$\gamma_q=$ 1&$\lambda_q$&4.14&2.36&1.49&1.56&1.61\\
&$n_g/B$&0.56&1.08&2.50&2.24&2.08\\
&$n_q/B$&3.11&3.51&5.16&4.81&4.62\\
$\gamma_g=$ 1&$n_{\bar q}/B$&0.11&0.51&2.16&1.81&1.62\\
&$n_{\bar s}/B$&0.05&0.11&0.25&0.22&0.21\\
&$\!P_{ch}\!$ [GeV/fm$^3$]&0.46&0.76&0.79&1.12&1.46\\
$\gamma_s=$ 0.15&$\rho_{\rm B}$&3.35&3.31&1.80&2.45&3.19\\
&$S/B$&12.3&19.7&41.8&37.4&34.9\\
\hline\hline
&$\gamma_s$&1&1&0.8&1&1\\
$\gamma_q=$ 1&$T_0$ [GeV]&0.184&0.215&0.233&0.239&0.255\\
&$\lambda_q$&4.14&2.36&1.49&1.56&1.61\\
$\gamma_g=$ 1&$n_g/B$&0.56&1.08&2.50&2.25&2.09\\
&$n_q/B$&3.11&3.51&5.12&4.81&4.60\\
&$n_{\bar q}/B$&0.11&0.51&2.12&1.81&1.62\\
$\gamma_s=$ 0.8&$n_{\bar s}/B$&0.34&0.68&1.27&1.43&1.33\\
or&$\!P_0\!$ [GeV/fm$^3$]&0.30&0.41&0.47&0.54&0.71\\
$\gamma_s=$ 1&$\rho_{\rm B}$&2.17&1.80&1.05&1.19&1.56\\
&$S/B$&14.5&24.0&49.5&46.5&43.4\\
\hline
\end{tabular} 
\end{center} 
\end{table}

After the collision has ended, for times $t\ge1$ fm/c, but
probably also $\le$ 3--5 fm/c, we relax the strange quarks to their
equilibrium abundance and the temperature drops from $T_{\rm ch}$
to the value $T_0$. 
The bottom portion of the table corresponds to this full chemical 
equilibrium (with exception of the S--W case for which we assume 
that strange quarks have reached 80\% of phase space occupancy 
as suggested by the experimental results \cite{analyze,eb}). 
During the formation of the strangeness flavor there is already 
evolution of the fireball outside of the collision region and 
we allow for this by keeping $\lambda_{\rm q}=$ Const. This effectively 
freezes the entropy content of gluons and light quarks,  allowing for 
significant drop in pressure and some cooling due to conversion 
of energy  into strangeness. 
 
In order to have some understanding of the possible thermal conditions 
prevailing in the early stages of the collision process, 
when the thermal equilibrium is reached, we have 
in the top section of the table \ref{bigtable}  selected some reasonable 
off-chemical equilibrium conditions --- we consider
20\% occupancy for gluons and light quarks, and 3\% for strange quarks and 
solve the same equations as for $T_{\rm ch}$, and the associated 
$\lambda_{\rm q}$\,. There is no change in 
the pressure, as the dynamical compression is present at 
this stage of the fireball evolution. But we see here in particular 
that the temperature $T_{\rm th}$ is considerably higher, 
since the number of quarks
and gluons present is considerably lower. 
 
The columns of table \ref{bigtable} correspond to the  cases of 
specific experimental interest, see top legend, in turn:   
Au--Au collisions at 
AGS, possible future Pb--Pb collisions at SPS with 40A GeV, 
S--Pb at 200A GeV, and for the Pb--Pb collisions at 158A GeV 
we considered two possible  values of stopping, see Eq.\,(\ref{Pdyn}): 
$\eta=0.75$ and  $\eta=1$\,. 
 
As discussed in section \ref{Init}, the temperature values shown in the 
bottom portion of the table are similar to the inverse slopes 
observed in particle spectra and shown in  Fig.\,\ref{energy}. 
Remarkably, the values of temperature $T_0$
found for the case of $E/B=8.6$ GeV at $\eta=0.5\pm0.1$ is 
just $233$ MeV, which corresponds nearly exactly to the reported 
inverse slopes of the WA85 results  \cite{WA85AIP},
and  $\lambda_{\rm q}=1.49$ also agrees exactly with the results of our 
analysis  \cite{analyze}, also shown in Fig.\,\ref{energy}. 
Even though there are  a number of tacit and explicit parameters 
(in particular $\eta=0.5, \alpha_s=0.6$) which enter this result the degree of 
the agreement is stunning and encourages us to explore in systematic 
fashion the variation of the key parameters with the energy content 
of the fireball.

\begin{figure}[tbh]
\vspace*{-0.5cm}
\centerline{\hspace*{-1.5cm}
\psfig{width=13cm,figure=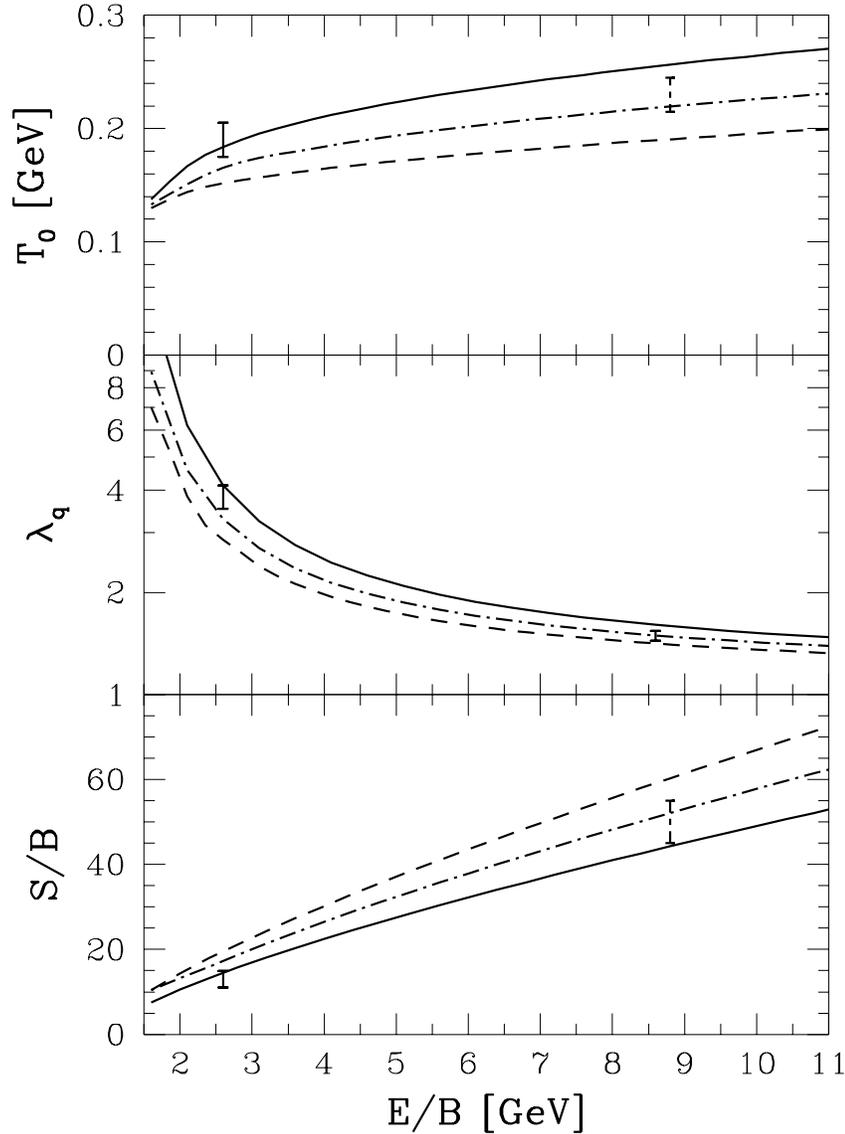}}
\vspace*{-0.5cm}
\caption{\protect\small 
Temperature $T_0$, light quark fugacity $\lambda_{\rm q}$
and entropy per baryon $S/B$ at the time of full chemical 
equilibration  as function of the QGP-fireball energy content 
$E/B$. Results for momentum stopping $\eta=1$ (solid line), 
0.5 (dot-dashed line) and 0.25 (dashed line) are shown. See text 
for comparison with analysis results.
\protect\label{fig1S95}
}
\end{figure}

In  Fig.\,\ref{fig1S95} we show, as
function of the specific energy content $E/B$, 
 the expected behavior of temperature $T_0$,  the
light quark fugacity $\lambda_{\rm q}$ and entropy per baryon
$S/B$ at the time of full chemical equilibration in the QGP fireball.  
The range of the possible values as function of $\eta$ is
indicated by showing  results, for $\eta=1$ (solid line), 0.5
(dot-dashed line) and 0.25 (dashed line). The experimental bars on
the right hand side  of the  Fig.\,\ref{fig1S95}
show for high (8.8 GeV) energy the result of analysis \cite{analyze} 
of the WA85 data \cite{WA85AIP}. The experimental bars  on
the left hand side of the  Fig.\,\ref{fig1S95} (2.6 GeV) are taken 
from our analysis of the BNL-AGS data  \cite{BNL-AGSthermal}, but note 
that in this case we had found $\lambda_{\rm s}=1.7$ and not 
$\lambda_{\rm s}=1$ as would be needed for the QGP interpretation 
at this low energy. For the BNL-AGS range of energies $E/B=2.3$--$2.6$ GeV,
 we expect $\eta \simeq 0.9$--1 and this is indeed  in  good agreement 
with the QGP-based evaluation  \cite{BNL-AGSthermal} of the experimental 
results ($T=180\pm30$ MeV, $\lambda_{\rm q}=4.8\pm0.4$ and $S/B=13\pm1$).  
We can explain in this analysis the case of S--S collisions which have a 
lot of flow \cite{analyzea,Heinzy}: we take at $E/B=9.6$ GeV a stopping 
fraction $\eta\simeq 0.3\pm0.1$ which yields  $T_0\simeq
196\pm13$ MeV, in agreement with the experimental results for 
the inverse slope \cite{NA35slope}. The value of $\lambda_{\rm q}$ 
which traces the  baryon density cannot be so easily estimated in 
this case, we refer to a study of the possible rapidity dependence 
of $\lambda_{\rm q}$ made recently  \cite{Heinzy}.
 
Among the key features in the  Fig.\,\ref{fig1S95}, we note
that, in qualitative terms, the drop in temperature with decreasing
energy and stopping is intuitively as expected, and the value of 
$\lambda_{\rm q}$ is relatively insensitive to the stopping power, 
and also varies little when the energy changes by $\pm$15\%. This
implies that even when different trigger conditions lead to 
different stopping fractions $\eta_i$, the resulting value of 
$\lambda_{\rm q}$ which is determining  the strange particle 
(baryon/antibaryon) ratios, is rather independent of different 
trigger conditions. Our analysis shows that {\it
$\lambda_{\rm q}$ decreases while $E/B$ increases}. This behavior can be
argued for by noting that baryon density is higher in the QGP at lower
energies. However, note that this intuitive insight was really arising
from our belief that the stopping of baryon number decreases as energy
of the collision increases, while the result here found occurs
irrespective of the change in baryon stopping.
This has considerable impact on
the  behavior of strange particle ratios as
function of collision energy (see section \ref{results}).
Another important result is the rapid rise
of specific entropy with the energy content: 
while at the BNL-AGS energies we 
find  similar entropy contents in the confined and
deconfined phases of hadronic matter, at CERN energies we encounter twice
as much entropy in the deconfined phase, which leads to a noticeable
excess in particle abundances  \cite{entropy}. 
We turn to discuss this important issue now.

\subsection{Entropy and Particle Multiplicity}
\begin{figure}[tb]
\vspace*{-7.6cm}
\centerline{\hspace*{0.7cm}
\psfig{width=15cm,figure=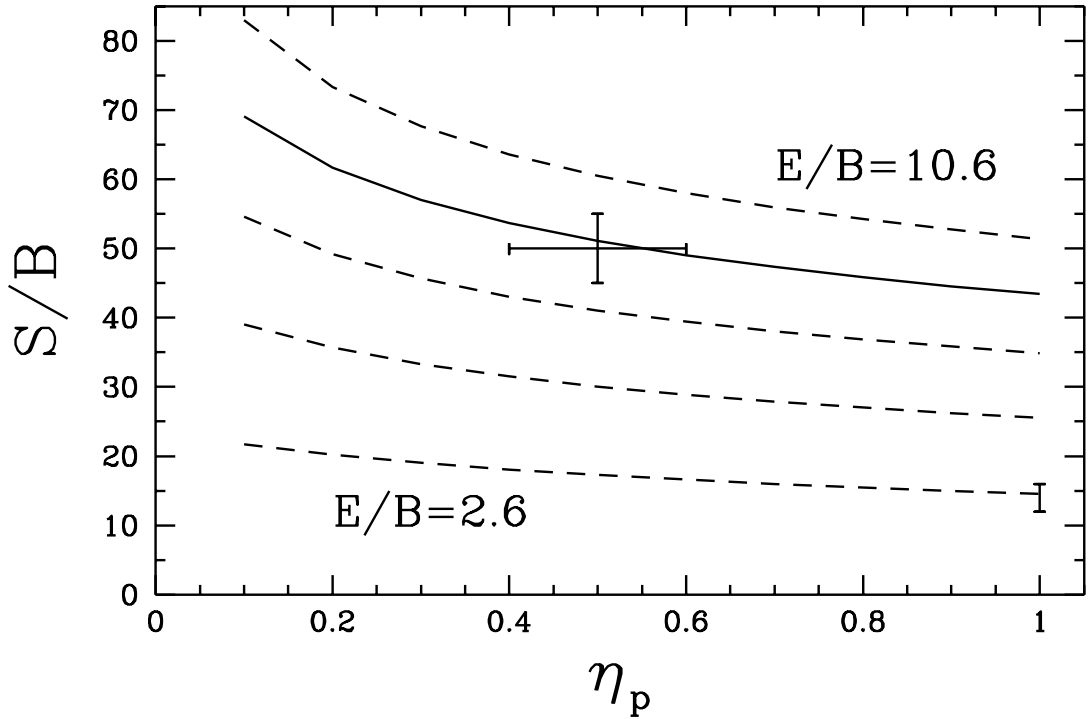}}
\vspace*{-3.6cm}
\caption{\protect\small 
Specific entropy $S/B$ as function of momentum stopping $\eta_{\rm p}$ at
given energy content of the QGP-fireball 
$E/B=2.6,\,4.6,\,6.6,\,8.6,\,$ and 10.6 
GeV. See text for discussion of the experimental points.
 \protect\label{entropy}}
\end{figure}

An important observable of the thermal fireball which allows 
to cross check the validity of our approach is its entropy content. 
We recall the first law of thermodynamics in the form:
\begin{equation}
E=TS-PV+\mu_B B\,,
\end{equation}
which with the relativistic EoS (negligible masses, no 
dimensioned scale in the interactions):
\begin{equation}
P={1\over 3}\left({E\over V}-4{\cal B}\right),
\end{equation}
leads to the simple relation:
\begin{equation}
{S\over B}={4\over 3} {1\over T} {E\over B} - \ln \lambda_{B}
-{4\over 3} {{{\cal B}V}\over {BT}}\,. \label{S1st}
\end{equation}
Note that this equation can only be applied to systems 
in chemical equilibrium. 
The two last terms in Eq.\,(\ref{S1st}) are comparatively 
small compared 
to the first term. We see that the entropy content can be 
relatively accurately 
estimated. For $E/B=8.8$ GeV and $T=235$ MeV we find 
$S/B=50$ in agreement with 
other studies \cite{analyze,entropy} and the 
results shown in the table \ref{bigtable} 
above. More complete numerical calculations yield the specific entropy at 
fixed specific energy as function of one statistical parameter, or as 
shown in  Fig.\,\ref{entropy}, as function of the momentum stopping. 
In  Fig.\,\ref{entropy} we show how the entropy per baryon
varies as function of stopping for different energies per baryon
$E/B=2.6$ to 10.6 GeV. The experimental bar in lower right corresponds to
our analysis of the BNL-AGS results \cite{BNL-AGSthermal}, and is in
agreement with stopping being greater than 80\% if QGP were formed at 
these energies. The cross in the middle of the figure is  
indicating the region in which we believe the S--W/Pb system to be 
(stopping $0.5\pm0.1$, 
specific entropy as discussed in \cite{analyze} and just above, see
Eq.\,(\ref{S1st}). We note that the solid line, corresponding to the 
energy content of S--W/Pb collisions, passes right through this region.

This fact that entropy per baryon is primarily dependent on the specific 
energy content in the fireball offers us the opportunity to explore 
further the hypothesis which had lead us to determine the relation between 
the kinematic energy and the fireball energy, see section \ref{energ}.
Should the energy stopping be greater than baryon stopping even by 15\%,
then the specific energy content would be much more greater, and in 
 Fig.\,\ref{entropy} we see that this would be implying a greater
momentum stopping $\eta_{\rm p} \to 1$. Since both energy and momentum 
stopping should increase together this is a possible scenario, and hence we 
must always remember that the 160--200A GeV collisions as possibly leading to 
a greater specific energy.  Conversely, if we were to assume that
energy stopping is smaller by 15\% than baryon stopping, this would imply 
a reduction of the specific energy and as seen in  
Fig.\,\ref{entropy} would require
$\eta_{\rm p} \to 0.2$, which is clearly not seen in the experiments.
We conclude that either all stopping fractions at CERN-SPS energies are
similar or that there is a somewhat smaller baryon stopping. The qualitative 
results about transverse energy production  \cite{stop} suggest that we 
should not consider $\eta_{\rm E}>0.5$ for 200A GeV  S--W/Pb collisions. 

\begin{figure}[tb]
\vspace*{-1.5cm}
\hspace*{-1.2cm}
\psfig{width=18cm,figure=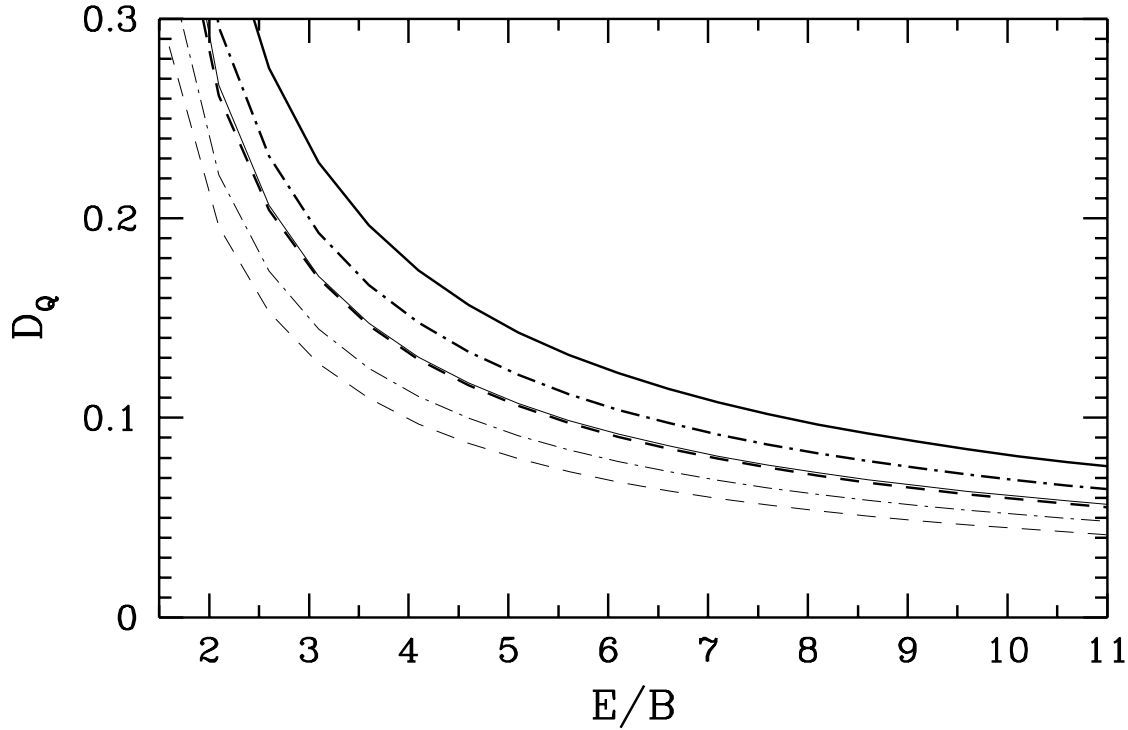}
\vspace*{-.8cm}
\caption{\protect\small 
$D_{\rm Q}$ versus $E/B$. Thick line $D_{\rm Q}\,S/B=4$, 
thin line $D_{\rm Q}\,S/B=3$.
$\eta=1 $ (solid line), $\eta=0.5 $ (dot-dashed line) and
$\eta=0.25 $ (dashed line).
 \protect\label{dq}}
\end{figure}

The experimentally measurable quantity is not the entropy, but the final 
state particle multiplicity  \cite{entropy,cool}. A useful 
observable is the quantity $D_{\rm Q}$:
\begin{equation} \label{DQdef}
D_{\rm Q}(y) \equiv {{\left({{dN^+\over dy} - {dN^- \over dy}}\right)}
\over {\left({{dN^+\over dy} + {dN^- \over dy}}\right)}}\,.
\end{equation}
Note that in the numerator of
$D_{\rm Q} $ the charge of particle pairs produced cancels out and it is
effectively a measure of the baryon number; thus in the product $D_{\rm
Q}$ with $S/B$ the baryon content cancels, and the result is roughly the
entropy content of the final state per final state pion number
(denominator of $D_{\rm Q}$). Specifically,  $D_{\rm Q}\cdot({S/B}) 
\simeq 3$ \cite{analyze} for a wide range of  $T,\lambda_{\rm q}$ and at 
$\lambda_{\rm s}=1$, and taking the 
population of heavy hadronic resonances emitted from the fireball 
as expected from a hadronic gas system in chemical equilibrium.  
However, the hypothesis of the 
equilibrated hadronic gas as a transient phase is highly inconsistent 
with the strange particle flow, which suggests no re-equilibration and 
direct disintegration of the fireball into final state hadrons. If this 
were the case, the production of heavy meson resonances would be 
suppressed against the thermal equilibrium expectations. Consequently, 
the dilution of the number of charged hadrons by decays of heavy neutrals 
would not occur and  we  should observe  $D_{\rm Q}^0\cdot({S/B}) 
\simeq 4.4$, where the superscript `0' indicates that we did not account 
for the particle decay in the quantity $D_{\rm Q}$ . 
In Fig.\,\ref{dq} we show as thick lines, using  $D_{\rm Q}\cdot({S/B})
\simeq 4$, how the observable $D_{\rm Q}$ depends on collision energy. 
Thin lines 
are for $D_{\rm Q}\cdot({S/B}) \simeq 3$ --- the momentum stopping varies 
between 1 (solid lines), 0.5 (dot dashed) and 0.25 (dashed). Experimentally, 
in S--Pb collisions the EMU05 collaboration finds $D_{\rm Q}=0.085\pm0.01$, 
shown at 8.8 GeV in Fig.\,\ref{dq}. As the CERN-SPS energy is lowered to 
4.3 GeV in CM frame (40A GeV  for the projectile) 
which is presumably accompanied by 
an increase in stopping to $\eta=1$, the value of  $D_{\rm Q}$ rises and 
reaches $D_{\rm Q}=0.16$, nearly twice the S--W/Pb results. We thus see 
that the widely different specific entropy values shown at the bottom 
of table \ref{bigtable} become measurable 
by means of the observable $D_{\rm Q}$ and conclude that $D_{\rm Q}$ is an extraordinarily valuable
experimental variable which differentiates structures arising in the
collision and from which in particular quantitative information about
specific entropy can be gained. 

We note in passing that a computation of the confined hadronic gas
entropy  \cite{analyze,entropy} at the same statistical conditions of the
fireball yields only 50\% of the QGP-fireball specific entropy value, 
and thus suggests in  view of the observed particle multiplicities that
the fireball could not have been just an assembly of confined hadrons.

\section{Strange Particles from QGP} \label{results}
\subsection{General Remarks}
The observable `strangeness' is more than just one  quantity which is
enhanced by a factor two or two and a half  when  one compares usual
$p$--$p$ and $p$--A with A--A reactions. The interesting aspects of 
this observable are that certain strange particles are much more 
enhanced than others, and moreover, the production mechanisms 
being very different from the usual ones, the behavior of the 
yields (cross sections)  with energy  can be expected to be 
of different nature.  A thermal fireball source can emit hadronic 
particles during its entire evolution. 
The observable strange particles are: K$^\pm$,  K$_{\rm S}$, $\phi$, 
$\Lambda$, $\overline{\Lambda}$,  $\Xi^-$, $\overline{\Xi^-}$, $ \Omega$ and
$ \overline{\Omega}$\,.  We should include here the closely related 
antiprotons  $\bar p$, which leads us to consider 11 types of particles. 
Leaving out an overall normalization factor associated with the reaction 
volume, and recalling that there is a constraint between the abundance
of kaons (K$^++$K$^-\simeq2$K$_S$) we have 9 independent normalization
parameters describing the yields. These can be redundantly measured with
the help of the $36=9\cdot 8/2$ independent particle yield ratios. Aside
of the yield normalization parameters, there are in principle 11
different spectral shapes  which we above implicitly presumed to be
related to each other: The experimental fact that once effects related to
particle decays and (transverse) flow are accounted for, the $m_\bot$
spectra are characterized by a common inverse slope parameter, cannot be
taken lightly and suggest strongly that the source of all strange
particles is indeed thermalized with a common temperature and that a 
common mechanism governs production of different strange particles.  

\subsection{Final State Particle Counting}\label{Counting}
It is easy to propose mechanisms
that would largely erase memory of the transient QGP phase. We will not 
discuss this class of {\it re-equilibrating} hadronisation pictures \cite{reequilib} of 
strange particles which seems not to be present at least at CERN-SPS
energies \cite{analyze}. 
Instead, we shall focus our attention on the alternative that
the particles emerge directly from the deconfined phase, be it because of
their radiative emission  \cite{RD87}, be it because a general rapid and 
explosive disintegration of the hot fireball occurs in the final stage 
of its evolution. 

The abundance of particles emerging in explosive disintegration 
or radiated is, according to Eq.\,(\ref{spectra}), determined by the 
normalization constant:
\begin{equation}\label{norm}
N_j=V\prod_i n_i\,,\qquad n_i=g_i\lambda_i \gamma_i ,
\end{equation}
where it is assumed that the final state particle of type $j$ contains 
the quark valence components of type $i$ and these are counted using 
their statistical degeneracy $g_i$, fugacity $\lambda_i=\exp(\mu_i/T)$ 
and the chemical equilibration factor $\gamma_i$. 
$ V$ is the emission source 
volume. Fragmentation of gluons could contribute to the abundance of  the
valence quarks and has been considered previously  \cite{RD87}. Because it
enhances the number of all quarks and the effect is weighted in a similar
way for all flavors, and further, since in the ratio of particle
abundances a partial cancelation of this effect occurs, this effect is
apparently of lesser importance.
 
Once the factor $\gamma_i$ accounts for the deviation from the full phase
space occupancy of the species $i$,  the chemical potentials for
particles and antiparticles are opposite to each other and the particle
and antiparticle  abundances are related, see Eq.\,(\ref{lam}).
As indicated in Eq.\,(\ref{norm}),  the fugacity of each final state 
hadronic species is the product of the valence quark fugacities. 
The abundances of the final state strange particles is gauged by
considering the Laplace transform of the phase space distribution
of strange particles, which leads to a partition function ${\cal Z}_{\rm s}$
like expression (\ref{4a}). 
The weight of individual components is
controlled by the non-equilibrium coefficients $\gamma_{\rm s}$, relative
meson and baryon abundance parameters $ C_{\rm B}^{\rm s}$, 
$ C_{\rm M}^{\rm s}$ and by the
fugacities $\lambda_{\rm q},\,\lambda_{\rm s}$:  
\begin{eqnarray}
\ln{\cal Z}_{\rm s} = { {V T^3} \over {2\pi^2} }
\left\{(\lambda_{\rm s} \lambda_{\rm q}^{-1} +
\lambda_{\rm s}^{-1} \lambda_{\rm q}) \gamma_{\rm s} C^{\rm s}_{\rm M}
F_K +(\lambda_{\rm s} \lambda_{\rm q}^{2} +
\lambda_{\rm s}^{-1} \lambda_{\rm q}^{-2}) \gamma_{\rm s} C^{\rm
s}_{\rm B} F_Y \right.\nonumber \\ 
\left.+ (\lambda_{\rm s}^2 \lambda_{\rm q} +
\lambda_{\rm s}^{-2} \lambda_{\rm q}^{-1}) \gamma_{\rm s}^2 C^{\rm
s}_{\rm B}  F_\Xi + (\lambda_{\rm s}^{3} + \lambda_{\rm s}^{-3})
\gamma_{\rm s}^3 C^{\rm s}_{\rm B} F_\Omega\right\}\ , 
\label{4a}
\end{eqnarray}
where the kaon, hyperon, cascade and omega 
degrees of freedom are included. Here $T$ is the freeze-out temperature.
The phase space factors $F_i$ of the strange particles are (with $g_i$
describing the statistical degeneracy):
\begin{eqnarray}
F_i&=&\sum_j g_{i_j} W(m_{i_j}/T)
\ .
\label{FSTR}
\end{eqnarray}
In the resonance sums $\sum_j$ all known strange hadrons should be
counted.  The function $W(x)$ arises from the phase-space integral of the
different particle distributions $f(\vec p)$. For the Boltzmann particle
phase space, when the integral includes the entire momentum range, the
well known result is found
\begin{equation}\label{therspec}
W(x)\equiv (4\pi)^{-1}\int d^3(p/T)f(\vec p)=x^2K_2(x)\,,
\end{equation}
where $x=m/T$ and $K_2(x)$ is the modified Bessel function.

Because the emission volume of the particles is not known, only the ratio
\begin{equation}\label{RsC}
R^{\rm s}_{\rm C}=C^{\rm s}_{\rm M}/C^{\rm s}_{\rm B}
\end{equation}  
appears in
observables of interest. Note that, only because we are allowing for
the possibility of the QGP formation, we must allow for the effect of
non-equilibrium hadronisation with $C_{\rm B,M}^{\rm s}\ne~1$
--- there is no reason
whatsoever to expect that the rapid disintegration of the deconfined
state will lead to particle abundances that are associated with chemical
equilibrium of the final state. We note that relative abundances of meson
and baryons emerging from hadronising QGP are difficult to equilibrate,
because processes which convert meson into baryon--antibaryon pairs are
relatively slow.
 
Determination of the freeze-out temperature and the non-equilibrium
hadronisation parameter $R^{\rm s}_{\rm C}$ requires that we observe the
kaon to (strange) baryon ratios. We will discuss this point below in 
section \ref{freezeout}\,.
 
\subsection{Strangeness Conservation}
There is a strong constraint between the two fugacities
$\lambda_{\rm q}$, and $\lambda_{\rm s}$ arising from the requirement of
strangeness conservation which was discussed at length recently 
\cite{analyze}. 
The manner in which this relation works,
depends critically on the nature of the particle source and we shall now
explore the meson-baryon off-equilibrium effects. 
These non-trivial relations between the parameters characterizing the final
state are in general difficult to satisfy and the resulting particle
distributions are constrained in a way which  differs considerably
between different reaction scenarios which we have considered in detail:
the rapidly disintegrating QGP or the equilibrated HG phase. These two
alternatives differ in particular by the value of the
strange quark chemical potential $\mu_{\rm s}$:
\begin{itemize}
\vskip -0.3cm 
\item[\it 1.] In a strangeness neutral QGP fireball $\mu_{\rm s}$
is always exactly zero, independent of the prevailing temperature and baryon
density, since both $s$ and $\bar s$ quarks have the same phase-space
size. 
\vskip -0.3cm 
\item[\it 2.] In any state consisting of locally confined hadronic
clusters, $\mu_{\rm s}$ is generally different from zero at finite baryon
density, in order to correct the asymmetry introduced in the phase-space
size by a finite baryon content.
\end{itemize}
At non-zero baryon density, that is for $\mu_{\rm B}\equiv 3\mu_{\rm
q}\ne 0$,  there is just one (or perhaps at most a few) special value
$\mu_{\rm B}^0(T)$ for which $\langle s \rangle = \langle \bar s \rangle$
at $\mu_{\rm s}^{\rm HG}=0$, which condition mimics the QGP. For the case
of a conventional  HG (Hagedorn type) we have studied these values
carefully \cite{Let94a}. For the final state described by 
Eq.\,(\ref{4a}), this condition of strangeness conservation takes the simple 
analytical form \cite{entropy,analyze}:
\begin{eqnarray}
\mu_{\rm q}^0=T{\cosh}^{-1}\left(R^{\rm s}_{\rm C}{F_{\rm K}\over
2F_{\rm Y}} -\gamma_{\rm s} {F_{\Xi}\over F_Y}\right),\quad \mbox{for}\
\mu_{\rm s}^{\rm HG}=0\ . \label{zero}
\end{eqnarray}
There is at most one non trivial real solution for monotonous arguments of $\cosh^{-1}$,
and only when this argument is greater than unity.   
 
Clearly, the observation  \cite{analyze} of $\lambda_{\rm s}=1\ 
(\mu_{\rm s}=0)$  is, in view of the accidental nature of this value 
in the confined phase, a
rather strong indication for the direct formation of final state hadrons 
from a deconfined phase, in which this is the natural value.  
We note that a further refinement  \cite{Heinzy} of  the original analysis 
 \cite{analyzea} of the S--S system at 200A GeV, which
allows for a rapidity dependence of $\lambda_{\rm q}$ due to flow 
further underpins the finding $\lambda_{\rm s}=0$.
We can thus safely conclude that strange particles 
produced in 200A GeV Sulphur interactions with diverse targets lead  to
a particle source which displays a symmetry in phase space size of 
strange and anti-strange particles. A natural explanation is that such  a
source is deconfined. It will be very interesting to see,
if this behavior is  confirmed for the Pb--Pb system and different
collision energies.
 
\subsection{Freeze-out Conditions}\label{freezeout}
This is not the place to develop a complete  hadronisation 
model of the QGP fireball --- instead we are trying to 
circumvent the need for a detailed model of hadronisation
 by introducing a global single parameter $R_{\rm C}^{\rm s}$ 
which will allow to determine the relative ratio of meson and 
baryon particle yields. There is a simple way to fix 
$R_{\rm C}^{\rm s}$: the strangeness conservation condition 
Eq.\,(\ref{zero}) must apply to the particle emitted from the 
fireball, with chemical conditions determined by the source: 
thus for each value of  $\mu_{\rm q}$ and $ T=T_{\rm f}$ there 
is a corresponding value of $R_{\rm C}^{\rm s}$ for which the 
strangeness remains balanced. $R_{\rm C}^{\rm s}= 1$ would almost 
always lead to asymmetric emission of the strange particles and 
thus in all cases to the formation of a final $\bar s$ or $s$ 
strangelett nuggets, except in  the unusual condition that the 
QGP disintegration occurs when the plasma and HG phase-spaces 
for $\bar s$ or $s$ particle-carriers are the same.  In any event, 
it is extremely instructive to study how the values of the so 
defined $R_{\rm C}^{\rm s}$ depend on the properties of the fireball. 
 
\begin{figure}[tbh]
\vspace*{-1.1cm}
\centerline{\hspace*{-.7cm}
\psfig{width=18cm,figure=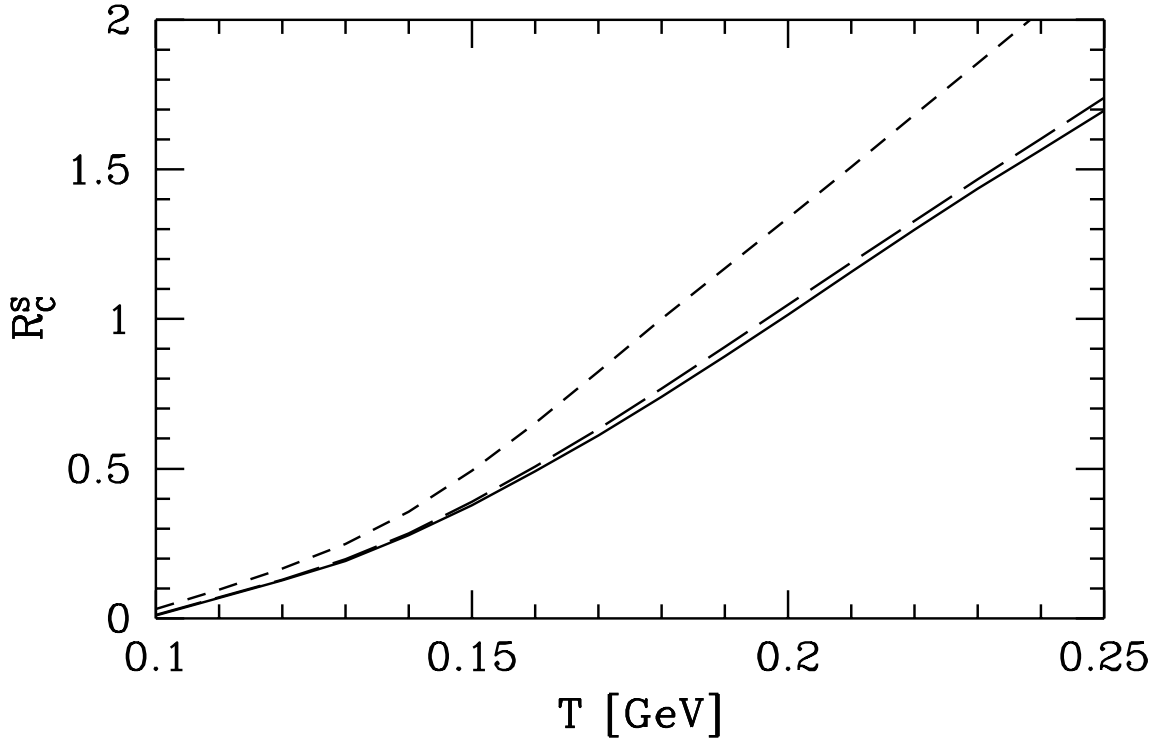}}
\vspace*{-.6cm}
\caption{\protect\small 
Strangeness neutrality line: $R_{\rm C}^{\rm s}$ versus freeze-out 
temperature for $\lambda_{\rm
q}=1.5$ (solid line), $\lambda_{\rm q}=1.6 $ (long dashed line) and 
$\lambda_{\rm q}=2.5$ (short dashed line).
 \protect\label{F2freeze}}
\end{figure}

We use here the experimentally motivated
$\gamma_{\rm s}=0.7$, though the deviation from unity is of little
numerical importance in present argument. Since the (multi-)strange
(anti)baryon particle ratios have led to $\lambda_{\rm q}=1.48$ and
$\lambda_{\rm s}=1.03$ we  use
the QGP freeze-out value $\lambda_{\rm s}=1$ and set $\lambda_{\rm q}$ to
three values in Fig.\,\ref{F2freeze}. The solid line is for $\lambda_{\rm
q}=1.5$, an appropriate choice for the case of S--W
collisions at 200A GeV, (when $\lambda_{\rm s}=1$). The short dashed
curve is for the choice $\lambda_{\rm q}=1.6$ which is as large as
$\lambda_{\rm q}$ will get for the case of Pb--Pb 160A GeV  collisions;
we see that we are finding rather narrow range of $R_{\rm C}^{\rm s}$ for
each given freeze-out temperature. The long dashed curve is for
$\lambda_{\rm q}=2.5$, the value which our model calculations 
suggest for the 40A GeV  collisions  (see table \ref{bigtable}).  
 
The value  $R_{\rm C}^{\rm s}=1$ is found for $T\simeq 200$ MeV at 
$\lambda_{\rm q}\simeq1.5$--$1.6$. For lower temperatures we have
$R_{\rm C}^{\rm s}<1$, which implies that the size of the strange 
baryon phase space needs to be increased compared to thermal equilibrium 
(or the strange meson phase space needs to be reduced) in order to 
allow a balance between strange and anti-strange quarks in emission 
from the QGP fireball.  
 
\begin{figure}[tbh]
\vspace*{-1.1cm}
\centerline{\hspace*{-1.2cm}
\psfig{width=18cm,figure=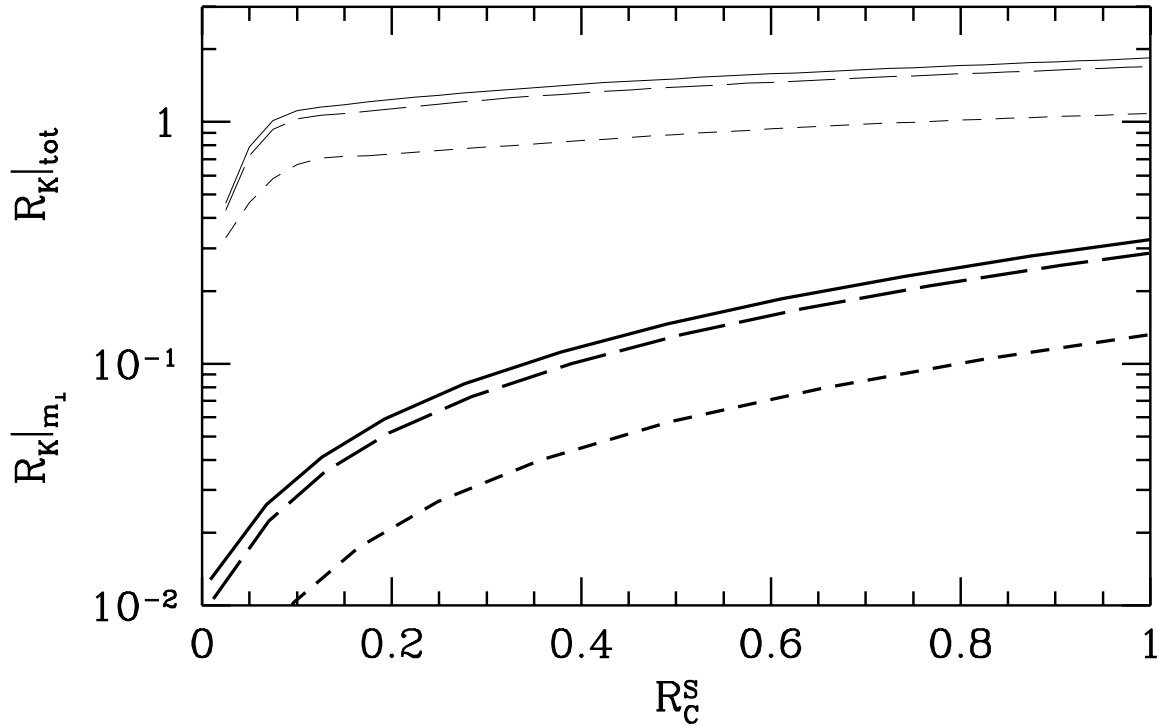}}
\vspace*{-.6cm}
\caption{\protect\small 
Thick lines: $R_{\rm K}\vert_{m_\bot}$, thin lines: 
$R_{\rm K}\vert_{\rm tot}$  
as function of $R_{\rm C}^{\rm s}$ for $\lambda_{\rm
q}=1.5$ (solid line), $\lambda_{\rm q}=1.6 $ (long dashed line) and 
$\lambda_{\rm q}=2.5$ (short dashed line).
\protect\label{Rcfreeze}}
\end{figure}

As noted in passing above, a description of  hadronisation 
using $R_{\rm C}^{\rm s}=1$ leads to a $s,\ \bar s$ asymmetry 
in the developing fireball. The question is which of
the two evolution options: the off-equilibrium particle emission, or 
strangeness distillation, is consistent with the experimental data. 
This can be decided considering a physical observable which is primarily 
sensitive to the parameter $R_{\rm C}^{\rm s}$, and to a lesser degree
to the other thermal model parameters. We choose to study  the kaon to
hyperon abundance ratio, both for all phase space and at fixed $m_\bot$. 
In the latter case we have: 
\begin{eqnarray}
    R_{\rm K}\vert_{m_\bot} \equiv\left.
    {K^0_{\rm s}\over \Lambda+\Sigma^0}\right|_{m_{\bot}} \simeq
     {R_{\rm C}^{\rm s}\over 8}\,{\lambda_{\rm s}/\lambda_{\rm d} +       
        \lambda_{\rm d}/\lambda_{\rm s} \over
               \lambda_{\rm s} \lambda_{\rm u} \lambda_{\rm d}} \, .   
\label{rk}
\end{eqnarray}
We made a preliminary study of this relation in \cite{Let94a}, 
which is valid when resonance decay contributions
cancel. To account here in necessary
detail for the resonance decay influence, we  have incorporated 
the decay pattern of all listed resonances numerically. The thick lines in 
Fig.\,\ref{Rcfreeze} give $R_{\rm K}\vert_{m_\bot}$ at fixed $m_\bot$
as function of $R_{\rm C}^{\rm s}$, see Eq.\,(\ref{rk}). 
We include the descendants of strong and weak decays in order to
facilitate comparison below with experimental data. We assumed that the 
distribution of parent particles for kaons 
and hyperons is according to the thermal equilibrium condition evaluated at 
temperature as given in the  Fig.\,\ref{F2freeze}. The values of 
$\lambda_{\rm q}=1.5,\,1.6,\,2.5$ and line conventions are  the same as 
used in that figure. The thin lines in  Fig.\,\ref{Rcfreeze} give results 
covering the full phase space. Since the kaon mass is much smaller than 
hyperon mass, this ratio  $R_{\rm K}\vert_{\rm tot}$ is considerably greater 
than  $R_{\rm K}\vert_{m_\bot}$, and hence we use logarithmic scale to 
display both  $R_{\rm K}$. We see in particular that in order to have 
HG-equilibrated yield of kaons and hyperons for fixed $m_\bot$, i.e. when 
$R_{\rm C}^{\rm s}=1$, we would have to find experimentally 
$R_{\rm K}\vert_{m_\bot}\simeq0.3$ in the S--W/Pb and/or Pb--Pb collisions. 

There is no officially reported value for the $R_{\rm K}$ ratio. However, 
collaboration WA85 has presented  \cite{WA85AIPdavis} a figure which shows 
for largely overlapping $1.1<m_\bot<2.6$ GeV the yields of 
$\Lambda,\,\overline{\Lambda}$  and $K_{\rm S}$ for the central
rapidity region $2.5<y<3$. 
No cascading corrections was applied to these results. Using graphic 
methods we obtained $R_{\rm K}\vert_{m_\bot}=0.11\pm0.02$. This implies 
a far off-HG-equilibrium result $R_{\rm C}^{\rm s}=0.4$ as shown
in   Fig.\,\ref{Rcfreeze}, and thus as seen in  Fig.\,\ref{F2freeze}
a freeze-out temperature $T_{\rm f}\simeq 150$ MeV. The possibility that 
$R_{\rm C}^{\rm s}\simeq 1$ ($R_{\rm K}\vert_{m_\bot}\simeq0.3$)
is experimentally  completely excluded. The factor $R_{\rm C}^{\rm s}\ne 1$ 
confirms the expectation that the reactions which change 
the number of baryons 
(baryon-antibaryon formation by  mesons) are relatively slow 
in the confined phase  \cite{KR85}, 
or that all strange particles originate directly from the QGP fireball. 
In any case we conclude that
the observed value $R_{\rm C}^{\rm s}=0.4$ allows to fix the
freeze-out conditions, here in particular $T_{\rm f}=150$~MeV.
 
The final issue is how, from the value  $R_{\rm C}^{\rm s}=0.4$, 
we can infer
the values of the abundance constants $C^{\rm s}_{\rm M}$ and
$C^{\rm s}_{\rm B}$ which (see Eqs.\,(\ref{4a},\,\ref{RsC})) express the 
relative strange meson and baryon production abundance to the thermal 
equilibrium values. If we argue that the strange meson abundance, akin 
to total meson abundance is enhanced by factor two (i.e. 
$C^{\rm s}_{\rm M}=2$) as we found studying the entropy enhancement, 
then the conclusion would be that the strange baryons are enhanced 
(against their tiny HG equilibrium abundance at $T_{\rm f}=150$ MeV) by 
factor $C^{\rm s}_{\rm B}=5$. It is clear from this observation that 
strange baryons and antibaryons are thus the key to the study of the
deconfined state. We now turn to study these particle yields.

\subsection{Strange Baryons and Antibaryons}\label{sbaab}
\begin{figure}[tb]
\vspace*{-0.5cm}
\centerline{\hspace*{-0.6cm}
\psfig{width=18cm,figure=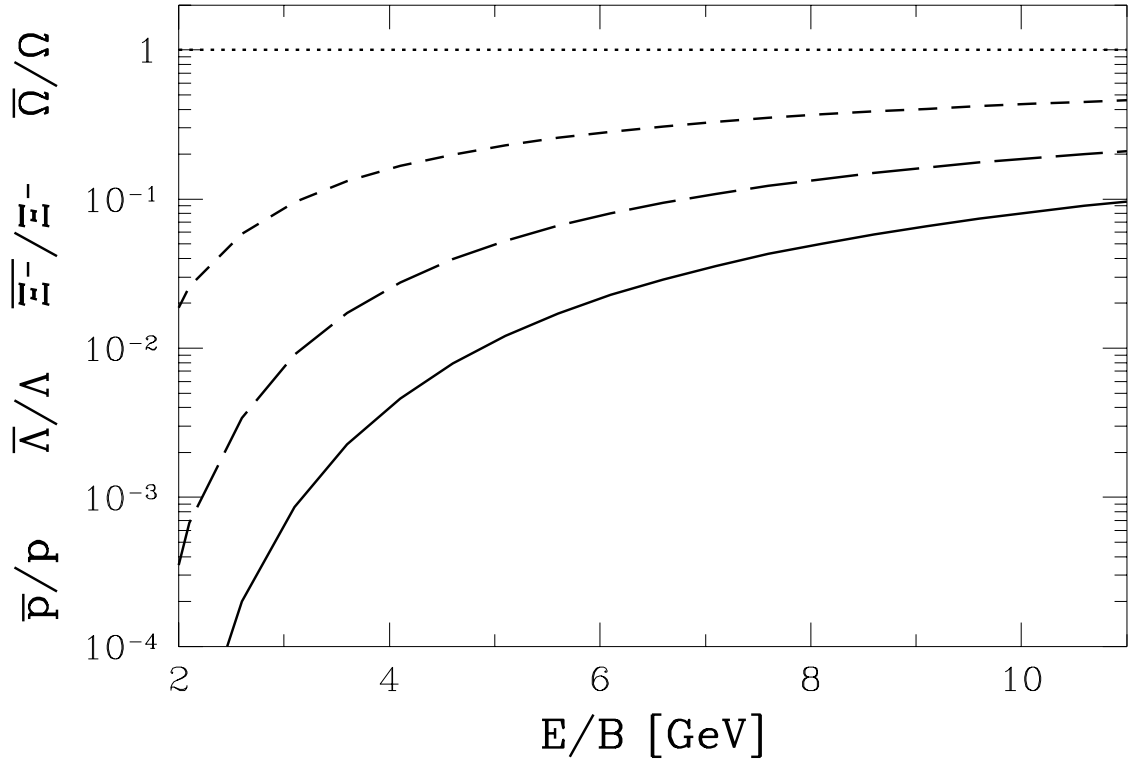}}
\vspace*{-0.6cm}
\caption{\protect\small 
Antibaryon to baryon abundance ratios as function of energy per 
baryon $E/B$ in a QGP-fireball: $R_{\rm N}=\bar p/p$ (solid line),
$R_\Lambda=\overline{\Lambda}/\Lambda$ (long-dashed line), 
$R_\Xi=\overline{\Xi}/\Xi$ (short-dashed line)
 and $R_\Omega=\overline{\Omega}/\Omega$  (dotted line)
\protect\label{eqratios}}
\end{figure}

The ratios of strange antibaryons to strange baryons {\it of same 
particle type\/}: $R_\Lambda=\overline{\Lambda}/\Lambda$, 
$R_\Xi=\overline{\Xi}/\Xi$ and $R_\Omega=\overline{\Omega}/\Omega$,
are in our approach simple functions of the quark 
fugacities and have been discussed at length recently 
 \cite{analyze,gammas}. In 
view of our current work we can predict the behavior of these
ratios as function of energy. Using the results for $\lambda_{\rm q}$ 
shown in  Fig.\,\ref{fig1S95}, we show in  Fig.\,\ref{eqratios} 
these three ratios. Since we assume  $\lambda_{\rm s}=1$\,, we have
$R_\Omega=\lambda_{\rm s}^{-6}=1$, but since  some re-equilibration 
is to be expected towards the HG behavior $\lambda_{\rm s}>1$, we expect
$\lambda_{\rm s}=1+\epsilon$, with $\epsilon$ small, 
and thus for this ratio $R_\Omega=1-6\epsilon<1$. 
A further non negligible correction which has been discussed at length 
in Ref.\,\cite{analyze} is due to the isospin asymmetry: in the
heavy Pb--Pb collisions it will be necessary to account for $d$--$u$ 
asymmetry which is as large as 15\%, and which favors the abundance of 
particles with $d$-quark content over those with $u$-quark content. 
This impacts here in particular the ratio $R_\Xi$, since there are 
no light quarks contributing to $R_\Omega$ and the ratio 
$R_\Lambda$ is $u$-$d$ symmetric. 

\begin{figure}[ptb]
\vspace*{2.5cm}
\centerline{\hspace*{2cm}
\psfig{width=15cm,figure=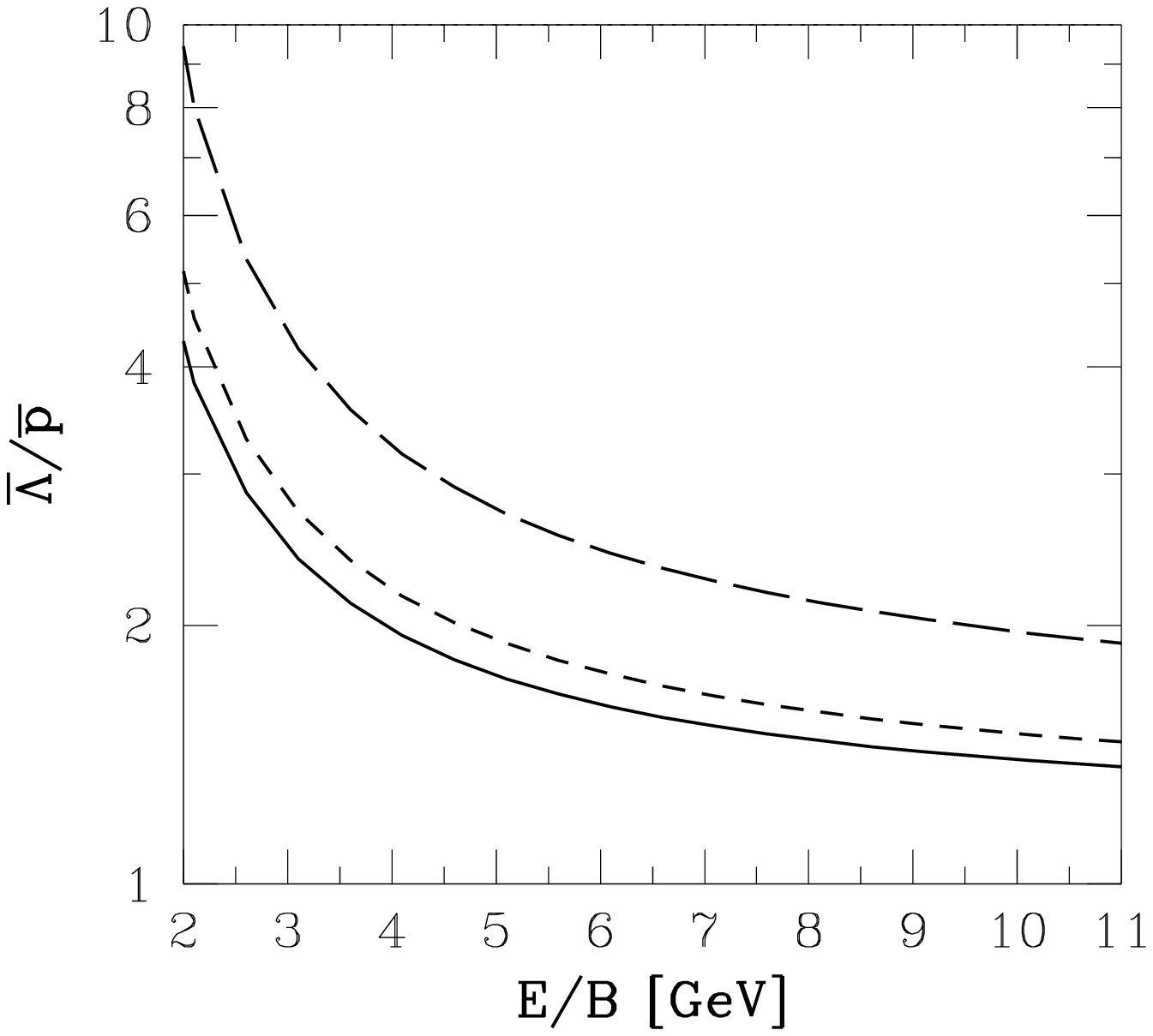}}
\vspace*{-3cm}
\caption{\protect\small 
Strange antibaryon ratio $\overline{\Lambda}/\overline{p}$,
as function of $E/B $ in a QGP-fireball;  
solid lines are for full phase space coverage, 
short dashed line  for particles with $p_\bot\ge 1$ GeV and
long dashed line for particles with $m_\bot \ge 1.7$ GeV.
 \protect\label{BARLP}}
\vspace*{3cm}
\centerline{\hspace*{2cm}
\psfig{width=15cm,figure=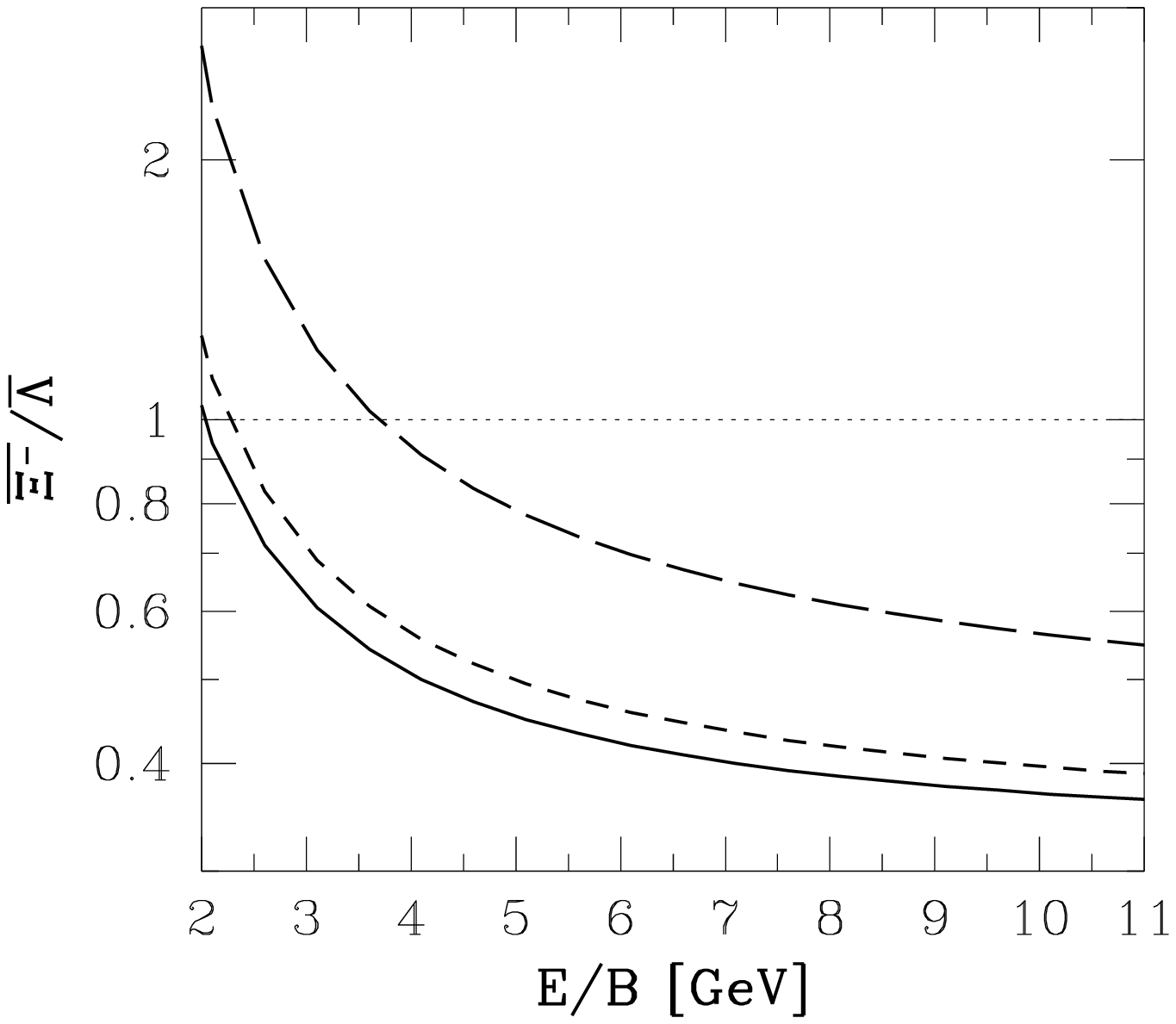}}
\vspace*{-3cm}
\caption{\protect\small 
Strange antibaryon ratio
$\overline{\Xi^-}/\overline{\Lambda}$  
with the same conventions as in Fig.\,\protect\ref{BARLP}.
 \protect\label{BARXL}}
\end{figure}

\begin{figure}[ptb]
\vspace*{2.5cm}
\centerline{\hspace*{2cm}
\psfig{width=15cm,figure=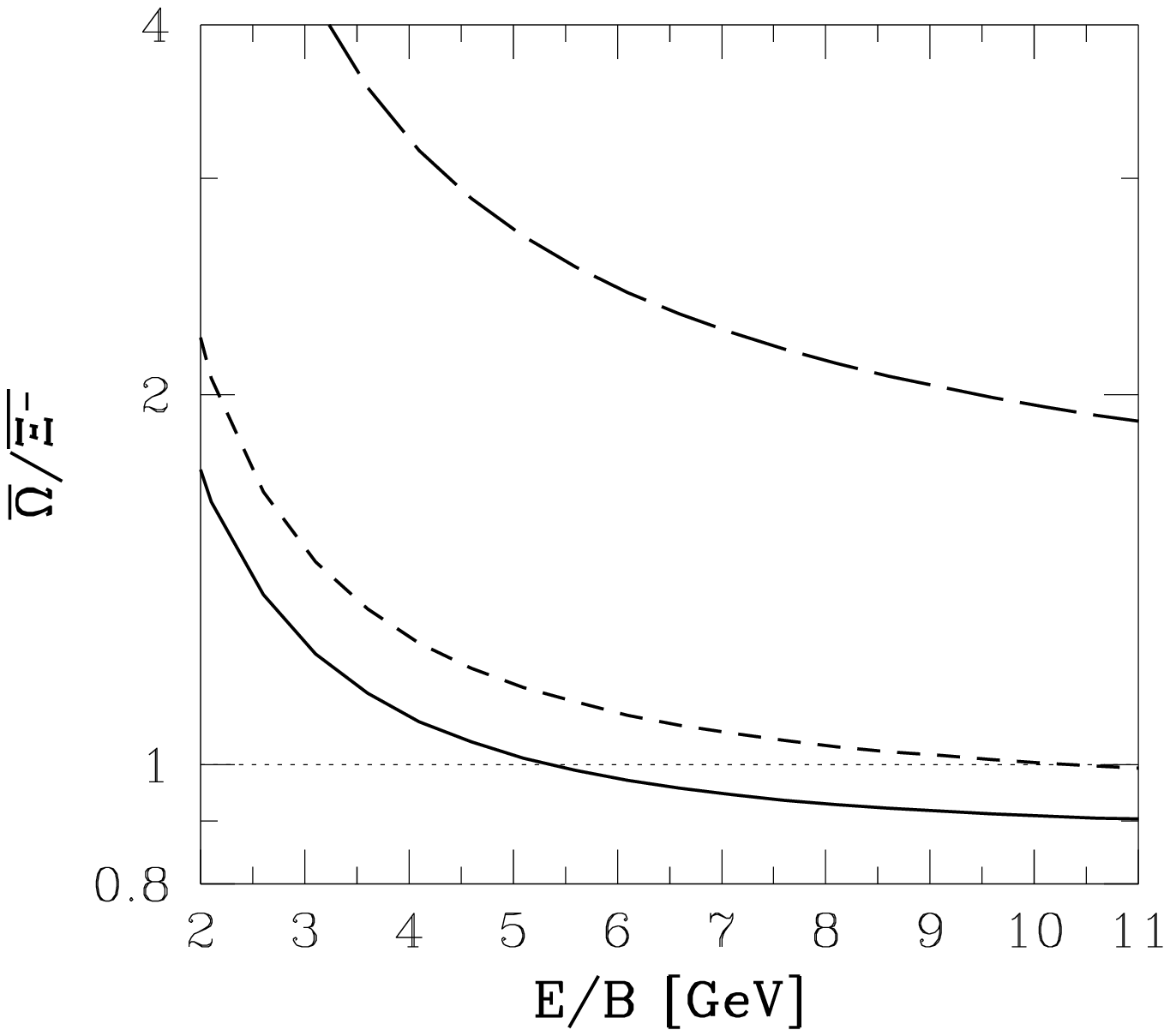}}
\vspace*{-3cm}
\caption{\protect\small 
Strange antibaryon ratio
$\overline{\Omega}/\overline{\Xi^-}$   
with the same conventions as in Fig.\,\protect\ref{BARLP}.
 \protect\label{BAROX}}
\vspace*{3cm}
\centerline{\hspace*{2cm}
\psfig{width=15cm,figure=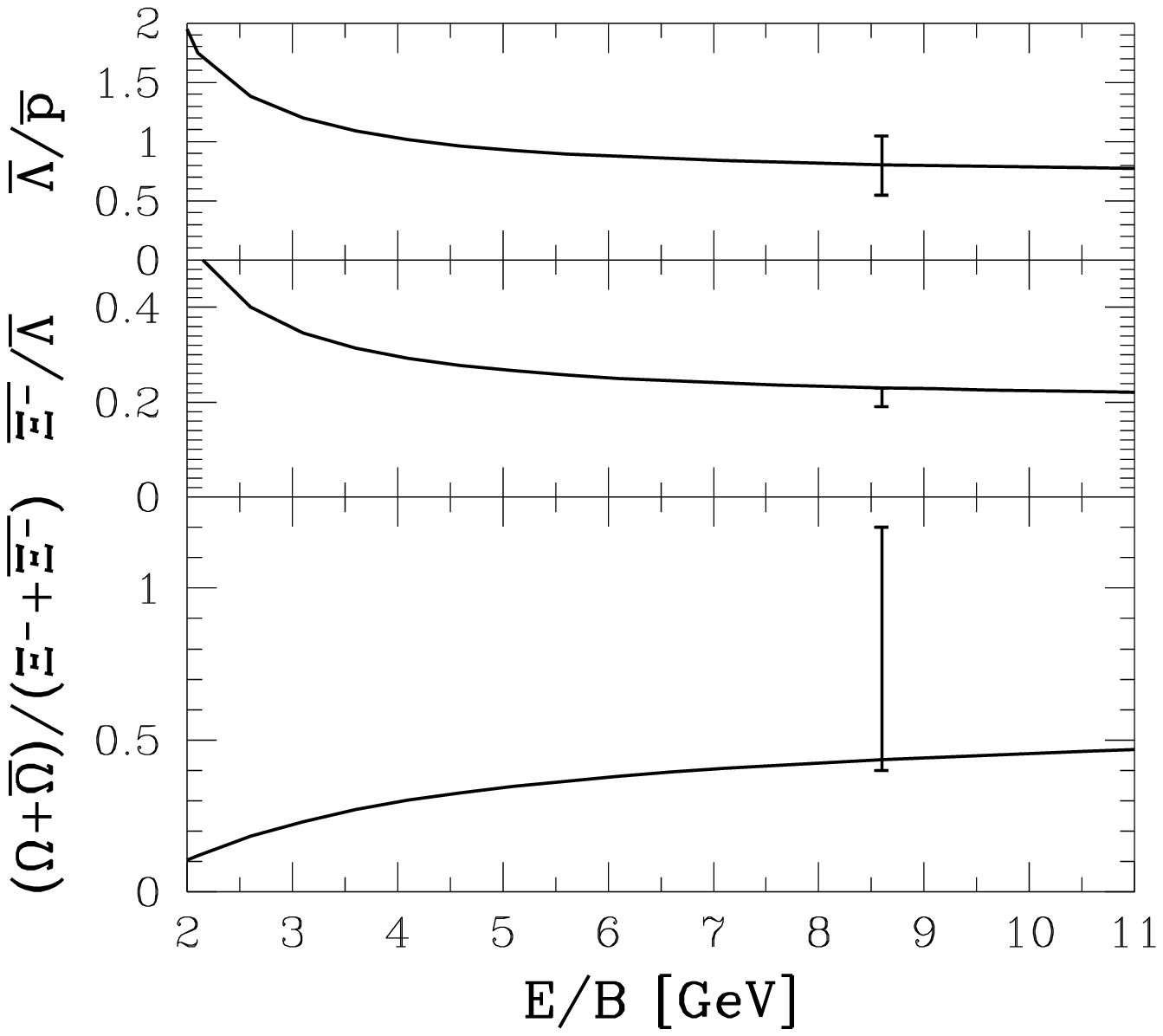}}
\vspace*{-3cm}
\caption{\protect\small 
Strange antibaryon ratios for S--W/Pb collisions as function 
of $E/B$ in a QGP-fireball:
$\overline{\Lambda}/\overline{p}$ (full phase space),
$\overline{\Xi^-}/\overline{\Lambda}$ for $p_\bot>1.2$ GeV and
$(\overline{\Omega}+\Omega)/(\overline{\Xi^-}+\Xi^-)$ for
$p_\bot>1.6$ GeV; experimental results 
shown are from experiments NA35, WA85.
 \protect\label{sratios}}
\end{figure}

We now study the ratios between antibaryons with different strange quark 
content. These are dependent on the degree of the strangeness saturation 
reached, and we shall take in this study $\gamma_{\rm s}=1$ assuming 
relatively large, long-lived system created in the collisions of largest 
available nuclei. In the Figs.\,\ref{BARLP}--\ref{BAROX} 
we show three ratios 
and for each ratio three results:  solid lines depicts the result for 
the full phase space coverage, short dashed line for particles with 
$p_\bot\ge 1$ GeV  and long dashed line for particles with $m_\bot 
\ge 1.7$ GeV. In  Fig.\,\ref{BARLP} we show the ratio 
$\overline{\Lambda}/\bar p$, in  Fig.\,\ref{BARXL} the ratio 
$\overline{\Xi^-}/\overline{\Lambda}$ and in  Fig.\,\ref{BAROX}  the ratio 
$\overline{\Omega}/\overline{\Xi^-}$. Because $\lambda_{\rm q}$ rises with
decreasing $E/B$, see   Fig.\,\ref{fig1S95}, we encounter overall 
the remarkable behavior that these three ratios increase as the collision 
energy is reduced. 

The behavior shown in Figs.\,\ref{BARLP}--\ref{BAROX} 
may be of considerable importance, if our expectations are confirmed 
that in reaction models in which QGP is not assumed and the particles 
are made in a sequence of microscopic collisions these ratios
 {\it do increase} with the collision energy, reflecting in this 
behavior the behavior of the reaction cross section observed in 
$p$--$p$ reactions. If this behavior is found at low energies and 
it matches at some energy at which a jump to the here presented yields, 
then the QGP behavior obtained here is truly
the smoking gun type evidence for the formation of the deconfined phase.
Furthermore, it would be a rather easy task to determine the transition
energy to QGP by merely seeking where these ratios peak as function of
$E/B$. 

As a final step in this discussion we present now the analysis of the 
available  and very recent WA85 $\Omega/\Xi^-$ production ratio  \cite{Omega} 
and the $\overline{\Lambda}/\bar p$ ratio
of the NA35 collaboration obtained  for the S--Au system at 200A GeV
 \cite{NA35pbar}.  Fig.\,\ref{sratios} shows a comparison of our ab initio
calculation and the pertinent experimental results. We use the same cuts
on the range of $p_\bot$ as in the experiment: the experimental points
show the results $\overline{\Lambda}/\bar p\simeq 0.8\pm 0.25$ 
(NA35) for full phase space,
$\overline{\Xi^-}/\overline{\Lambda}=0.21\pm0.02$ (WA85) for $p_\bot>1.2$
GeV; and $(\Omega+\overline{\Omega})/(\Xi^-+\overline{\Xi^-})=0.8\pm0.4$
(WA85) for $p_\bot>1.6$ GeV. The  chosen value of $\gamma_{\rm s}=0.70 $
and $\eta_{\rm p}=0.5$  brings about good agreement of our model 
with the precise value of $\overline{\Xi^-}/\overline{\Lambda}$. 
  Fig.\,\ref{sratios} shows also the impact of the change of the
collision energy on these results, using 50\% stopping, rather than 
$\eta=1$ used in  Figs.\,\ref{BARLP}--\ref{BAROX}. 

Considering that we
have computed here everything in an ab initio  dynamical model (which 
as discussed above has some tacit and explicit parameters such as the
QCD coupling $\alpha_{\rm s}=0.6$ etc., chosen to be in agreement with 
the earlier experimental results) it is remarkable that such a 
good agreement with the two very recent results could be attained.
We can conclude that the fact that the  two ratio
$\overline{\Lambda}/\bar p$ (NA35) and 
$(\Omega+\overline{\Omega})/(\Xi^-+\overline{\Xi^-})$ (WA85)
are satisfactorily  explained, provides a very  nice confirmation of 
the consistency of the thermal fireball model. 

\begin{figure}[tb]
\vspace*{-0.5cm}
\hspace*{-1.1cm}
\psfig{width=18cm,figure=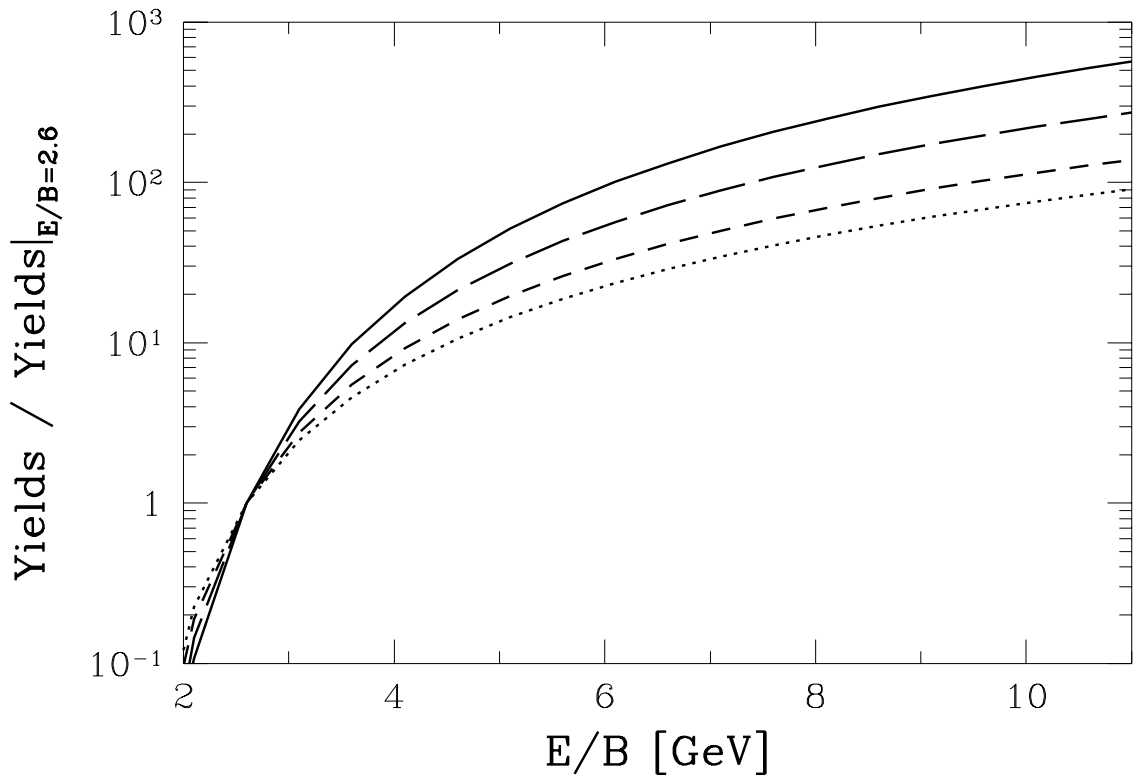}
\vspace*{-0.6cm}
\caption{\protect\small 
Relative antibaryon yields as function of $E/B$ in a QGP-fireball.
$\overline{p}$ (solid line), $\overline{\Lambda}$  (long-dashed line)
$\overline{\Xi^-}$ (short-dashed line) and $\overline{\Omega}$ (dotted
line), all normalized to their respective yields at $E/B=2.6$ GeV\,.
\protect\label{PLYIELDSNORM}}
\end{figure}

An interesting question which arises quite often is how the 
particle and in particular antibaryon yields vary with energy. 
Eq.\,(\ref{4a}) allows to determine the absolute particle 
yields as function of fireball energy. Considerable uncertainty
is arising from the off-equilibrium nature of the hadronisation 
process, which in particular makes it hard to estimate how the 
different heavy particle resonances are populated, and also, 
how the abundance factors $C_{\rm B}^{\rm s}$ vary as function 
of energy. Some of this uncertainties are eliminated when we 
normalize the yields at an energy, which we take here to be 
the value $E/B=2.6$ GeV which is applicable to the 
BNL-AGS experiments. In  Fig.\,\ref{PLYIELDSNORM} 
the so normalized yields of antibaryons  taking the freeze-out 
temperature $T=150$ MeV are shown (we also assume $\gamma_{\rm s}=1,\, 
\eta_{\rm p}=1$ and absence of any re-equilibration after particle
emission/production). These yields are rising in qualitatively 
similar systematic fashion with energy, as would be expected from the 
microscopic considerations, but the rise of more strange antibaryons 
is less pronounced, unlike what we would naively have expected.
The quantitative point to note is that at BNL-AGS 
($E/B=2.6$ GeV) the yield from a disintegrating QGP-fireball is a 
factor 100--400 smaller compared to yields at $E/B=$9 GeV. Since the 
particle density $dN/dy$ is not that much smaller at the lower energies 
(recall that the specific entropy, see table \ref{bigtable}\,,
drops only by factor 3.5, implying a reduction in specific multiplicity 
by a factor 5), it is considerably more difficult at the lower energies 
to search for antibaryons than it is at higher energies. We should 
remember that the results presented in Fig.\,\ref{PLYIELDSNORM} 
are obtained assuming formation of the QGP-fireball and same 
freeze-out and hadronisation conditions for all energies shown.

\section{Summary and Conclusions}
The key issue of interest to us is the identification of the 
deconfined quark-gluon plasma state in  relativistic nuclear 
collisions. We have shown \cite{analyze} in recent work that many features of 
strange particle production results obtained at 200A GeV, are consistent with 
the QGP hypothesis; and we have argued that the conventional reaction 
picture involving a fireball made of confined  hadrons is for a 
number of reasons incompatible with these experimental results. 
We felt that in order to ascertain the possibility that indeed the 
QGP phase is already formed at 200A GeV a more systematic exploration 
as function of collision energy of these observables would be needed ---  
 conclusions drawn from one set of experimental conditions 
suffer from the possibility that some coincidental and unknown 
features in the reaction mechanisms could lead to the observed 
properties. It is highly unlikely that this would remain the case,
should a key feature of the experiment be varied. 
 
Variation of the number of participating nucleons   has in 
the QGP reaction picture the considerable disadvantage that 
we reduce the size of the deconfined region, which even 
in the case of Pb--Pb reactions, is not very large. This
very likely will reduce the lifespan of the deconfined 
state, and the key reaction measure, the stopping 
`friction' is reduced in such a case as well. All these effects make an 
interpretation of the data with varying the number of participants, 
in terms of the same reaction model, exceedingly  difficult, 
if not impossible. Against this drawback, the study of the particle 
production with change in the reaction energy  (measurement of 
excitation functions) appears to be the best approach, which 
preserves the maximum attainable size and lifespan of the hypothetical
deconfined region, while minimizing the change in the stopping, in 
particular if one is operating in the energy domain in which for the 
largest available nuclei the stopping is nearly complete. Furthermore, 
at some low energy, we must reach the conditions of the normal hadron 
cascade interaction, and thus be able to describe the results reliably 
in terms of established cross sections.

Thus our primary objective in this work was to establish the systematic 
behavior of the antibaryon yields as the energy and stopping of the 
colliding large nuclei vary, and to determine how the freeze-out 
conditions  of these particles can be determined. This is necessary since 
the experimental results suggest that while the thermal (kinetic) 
equilibrium is established, the chemical (particle abundance)
equilibrium in the processes governing final state particle freeze-out
is largely not achieved. We thus developed a model which relates the energy 
content of the fireball to the collision energy and the stopping friction. 
In order to relate the energy to the statistical properties of the fireball 
which in our approach controls the particle yields we had to use EoS 
 of the deconfined matter, and in particular to account for the 
impact of QCD-interactions on the properties of the QGP, see 
section \ref{EOSQGP}. We have then shown 
how the study of kaon to hyperon ratios allows to determine the meson--baryon
chemical non-equilibrium parameter. We found that while the entropy production
suggests a meson yield which is about twice that of thermal freeze-out yield,
the baryon abundance is five times greater than the small thermal freeze-out 
abundance.

We believe that the thermal equilibrium approach should not be 
advanced in the context of A--A collisions as being just an 
economical, but otherwise approximate and highly limited 
method. Actually, one can argue that this is a more appropriate approach 
compared to microscopic dynamical (cascade) models, pending a better 
understanding of the elementary soft particle and entropy producing 
 \cite{Divonne}  hadronic processes. 
In the A--A  reactions there is no reason to expect that  `cascades' of
$p$--$p$ type interactions, which are, as matter of principle, even less 
understood, lead to adequate understanding of soft particles in nuclear 
collisions. 

We have used a simple  model which allows to determine, in a systematic 
fashion, the thermal conditions reached in high density deconfined matter 
generated in heavy ion collisions. It is based on the observation that
during the collision the compression of the quark-gluon matter can proceed 
until the internal pressure succeeds in stopping further impact 
compression. In this picture we worked far from chemical quark-gluon
equilibrium, but always assumed that the thermal equilibrium is 
reached rapidly, even on the scale of the nuclear collision time, 
which is believed to be $\simeq 1$ fm/c given the geometric size of 
the nuclei and the Lorenz contraction factors.

Motivated by the absence of chemical particle abundance equilibrium,
natural for a thermal HG in the final state, we employ a picture of 
particle production which involves rapid disintegration of the 
QGP-fireball\ \cite{gammas}. Central to the particle abundances are then the 
chemical properties of the QGP-fireball and we have discussed these 
comprehensively as function of collision energy and stopping 
in section \ref{finals}. We have shown that the thermal conditions we
find at the end of strangeness chemical equilibration in the fireball,
see bottom of table \ref{bigtable}\,,
are in good agreement  with our expectations derived from
particle yields seen in S--Pb/W collisions. To carry through this program
we needed to make a reasonable choice of remaining parameters:
in particular stopping $\eta=50$\% --- about equal for baryon number,
energy and pressure; perturbative EoS of particular
form we discussed at length in section \ref{EOSQGP}. Given these 
assumptions and parameters, we were able to study, 
 the current strange particle data at 200A GeV and
have reached a very satisfactory agreement with experiment as is shown in
Fig.\,\ref{BARLP}.
  
With this encouragement, we  have computed in a systematic 
fashion the behavior of strange particle (kaon, baryon and 
antibaryon) yields assuming conditions likely to occur in Pb--Pb 
interactions (e.g. full
stopping and $\gamma_{\rm s}=1$). It is most interesting that these results 
show patterns of behavior which could
indeed be unique for the QGP type of fireballs --- in particular, the
relative yields of strange antibaryons 
(see Figs.\,\ref{BARLP} -- \ref{BAROX}) lead to greater ratios occurring 
at smaller collision energies, down to very low energies, where in models 
involving  cascading type reactions such ratios would be so small that 
their measurement would be difficult. We are persuaded 
that this pattern of behavior could not
occur for normal confined matter, where the rise in cross sections with 
energy dominates particle yields whenever these are not arising from 
collective phenomena such as is a deconfined QGP phase. 

A much discussed question \cite{Cley93,CR93,Let94a} has been whether
current strange particle data are consistent with the fully equilibrated
hadronic gas (HG) picture of hot hadronic matter. The motivation for such
a hypothesis emanates from the observed rapid thermalization: given
that we do not understand the mechanisms of thermalization, one could
equally argue for the `a priori' presence of fully equilibrated (thermal
and chemical) HG phase. We do not believe that the chemical and thermal
equilibration are due to same physical processes, and thus while
allowing thermal equilibrium we maintain the possibility in our analysis
that chemical equilibrium develops slowly and not completely 
during the reaction. 
The results presented here support this point of view.

We stress that our description of particle production is based on
collective mechanisms (QGP-fireball) and is thus intrinsically 
different from  microscopic approaches, in
particular when these are based on a hadronic cascade picture. Such
models generally exploit specific data and/or extrapolations and
assumptions about individual hadronic reactions and their cross sections. 
If the true underlying thermalization processes are different (as
is likely) from those used in current microscopic approaches, and are
indeed much faster than cascades suggest, the whole representation  of
the collision evolution in these microscopic approaches needs to be 
reviewed --- in this perspective our thermal approach, already 
advocated for the $p$--$p$ reactions  \cite{HRblack} maybe seen as 
an experimentally well motivated hypothesis. 
We note that no alternative model to the here developed rapidly
hadronising QGP has been proposed which could generate both strangeness
abundance and  multi-strange antibaryon enhancement. For example the
description in terms of the dual parton model DPM, \cite{Cap95}, which
introduces a number of parameters to enhance strangeness and 
strange antibaryons, arrives
at a considerably smaller relative abundance $\overline{\Omega}
/\overline{\Xi}$\,.

Some of our results, though presented 
in great numerical detail, could suffer 
from considerable systematic errors, in particular when  
overall absolute yield normalization is discussed. For example, 
to determine the freeze-out conditions we need to determine the 
kaon yield after all heavy resonances have decayed. We employed 
all known tabulated resonances (and hence probably not all), and 
we adopted the statistical spin/isospin factors not always
established. Furthermore, there is the influence 
of the large resonance width which allows that these states are formed  
at energy below the resonance mass. There is also the possibility 
that some resonances alter their properties (particle width, mass) 
in dense matter. All these effects may have significant 
impact on the relation we have established between kaon to hyperon ratio 
and thus on the freeze-out conditions, and off-equilibrium properties, which 
in turn impacts the individual  particle yield. The presence
of these effects and uncertainties is also behind our omission from 
this discussion of the $\phi(\bar s s)$-meson, which participates 
and benefits from the general strangeness enhancement effects, but 
which is difficult to describe with the precision that is needed. 
 
Fortunately, these systematic uncertainties have very small impact on  
the strange antibaryon ratios, in particular when these are considered 
at fixed, high $m_\bot$, and we firmly believe, in view of the results
we have obtained here, that such data provide the best hadronic 
signatures, and diagnostic tools, of the deconfined matter. We recall
 the large ratios in the QGP-fireball reaction picture,
such as $\overline{\Xi}/
\overline{\Lambda}$ which we have found at relatively small energies ---
in microscopic models and near to 
$\overline{\Xi}$ production threshold in $p$--$p$ interaction 
this ratio is very small. This lets us 
expect that there will be a peak in the relative $\overline{\Xi}/
\overline{\Lambda}$ yield as function of 
collision energy which will  provide an interesting possibility to 
identify the energy at which collective production of strange antibaryons
is first encountered. At this energy we should also encounter for the 
first time the other features of the QGP phase: strangeness production 
enhancement, strange phase space saturation ($\gamma_{\rm s}\to 1$)\,, 
entropy enhancement (particle multiplicity enhancement), pattern of 
strange antibaryon flow showing $\lambda_{\rm s}=1$. 

\vspace{0.5cm}
\subsection*{Acknowledgment}
 J.R. acknowledges partial support by  DOE, grant
		DE-FG03-95ER40937 \,. 


\end{document}